\documentclass[letterpaper,twocolumn,10pt]{article}
\usepackage{usenix2019_v3}

\usepackage{tikz}
\usepackage{amsmath}

\usepackage{filecontents}

\usepackage{url}
\usepackage{wrapfig}
\usepackage{bbding}
\usepackage{graphicx}
\usepackage[ruled,vlined,linesnumbered]{algorithm2e}
\usepackage[x11names]{xcolor}
\usepackage{listings}
\usepackage{algpseudocode}
\usepackage{xcolor}
\usepackage{tikz}
\definecolor{javared}{rgb}{0.6,0,0} 
\definecolor{javagreen}{rgb}{0.25,0.5,0.35} 
\definecolor{javapurple}{rgb}{0.5,0,0.35} 
\definecolor{javadocblue}{rgb}{0.25,0.35,0.75} 
\definecolor{DarkGreen}{rgb}{0.0, 0.5, 0.0} 
\usepackage{enumitem}
\SetKwInput{Input}{Input}
\SetKwInput{Output}{Output}
\usepackage{subcaption}
\usepackage{colortbl}
\usepackage{multirow}

\lstdefinestyle{mystyle}{
      language=C++,
        basicstyle=\ttfamily\scriptsize, 
        keywordstyle=\color{javapurple}\bfseries,
        stringstyle=\color{javared},
        commentstyle=\color{javadocblue},
        morecomment=[s][\color{javadocblue}]{/**}{*/},
        numbers=left,
        breaklines=true,
        numberstyle=\tiny\color{black},
        stepnumber=1,
        numbersep=5pt,
        tabsize=2,
        showspaces=false,
        showstringspaces=false,
        morekeywords={foreach, uint64_t, uint16_t},
        classoffset=0,
        xleftmargin=1.8em,
        escapeinside={(*@}{@*)},
        moredelim=*[is][\color{red}]{[[[}{]]]},
        captionpos=b,
}

\usepackage{titlesec}
\titleformat{\subsubsection}[runin] 
  {\normalfont\bfseries} 
  {\thesubsubsection} 
  {} 
  {} 

\newcommand{\name}[0]{Hummingbird\xspace}

\newcommand{\us}[0]{{\textmu}s\xspace}

\newcommand{\upth}[0]{\textsuperscript{th}\xspace}

\newcommand{\codeIn}[1]{{\small\texttt{#1}}}

\newcommand{\eg}{\hbox{\emph{e.g.}}\xspace}
\newcommand{\ie}{\hbox{\emph{i.e.}}\xspace}

\newcommand{\MyPara}[1]{\noindent\textbf{\textit{#1}}~}

\definecolor{ForestGreen}{RGB}{34,139,34}

\newcommand{\htc}[1]{\textcolor{black}{#1}}

\begin{document}
\title{\textbf{\name: SLO-Oriented GPU Preemption at
Microsecond-scale
}}



\author{
  Tiancheng Hu\textsuperscript{12},
  Chenxi Wang\textsuperscript{3*},
  Ting Cao\textsuperscript{4},
  Jin Qin\textsuperscript{3},
  Lei Chen\textsuperscript{3},
  Xinyu Xiao\textsuperscript{5},
  Junhao Hu\textsuperscript{12},\\
  Hongliang Tian\textsuperscript{6},
  Shoumeng Yan\textsuperscript{6},
  Huimin Cui\textsuperscript{3},
  Quan Chen\textsuperscript{7},
  Tao Xie\textsuperscript{12*}\\
  \textsuperscript{1}SCS, Peking University, Beijing, China\\ 
  \textsuperscript{2}Key Lab of HCST (PKU), MOE, Beijing, China\\
  \textsuperscript{3}University of Chinese Academy of Sciences, Beijing, China\\ 
  \textsuperscript{4}Institute for AI Industry Research, Tsinghua University, Beijing, China \\
  \textsuperscript{5}Huazhong University of Science and Technology \\
  \textsuperscript{6}Ant Group \\
  \textsuperscript{7}Shanghai Jiao Tong University
}


\maketitle
\renewcommand{\thefootnote}{\fnsymbol{footnote}}
\footnotetext[1]{Corresponding authors: Chenxi Wang and Tao Xie.}
\begin{abstract}

Existing GPU-sharing techniques, including spatial and temporal sharing, aim to improve utilization but face challenges in simultaneously ensuring SLO adherence and maximizing efficiency due to the lack of fine-grained task scheduling on closed-source GPUs.
This paper presents \name, an SLO-oriented GPU scheduling system that overcomes these challenges by enabling microsecond-scale preemption on closed-source GPUs while effectively harvesting idle GPU time slices. Comprehensive evaluations across diverse GPU architectures reveal that \name improves the SLO attainment of high-priority tasks by 9.7$\times$ and 3.5$\times$ compared to the state-of-the-art spatial and temporal-sharing approaches. When compared to executing exclusively, the SLO attainment of the high-priority task, collocating with low-priority tasks on \name, only drops by less than 1\%. Meanwhile, the throughput of the low-priority task outperforms the state-of-the-art temporal-sharing approaches by 2.4$\times$.
\name demonstrates significant effectiveness in ensuring the SLO while enhancing GPU utilization.
\end{abstract}

\pagestyle{plain}
\section{Introduction}
In recent years, deep neural network (DNN) models, particularly Transformer-based architectures such as ChatGPT~\cite{chatgpt} and Gemini~\cite{gemini}, have become increasingly prevalent. Their escalating computational demands for both online inference and offline training have placed unprecedented pressure on available GPU resources. However, the deployed GPUs are usually allocated in a coarse-grained scheme that dedicates several GPUs for a specific task to guarantee the service-level objective (SLO), resulting in extremely low GPU utilization~\cite{gpu-workload-analysis@atc19,mlaas@nsdi22,hived@osdi20}. For example, GPU utilization is only about 52\% in Microsoft's GPU cluster \cite{gpu-workload-analysis@atc19} and even less than 25\% in Alibaba's GPU cluster \cite{mlaas@nsdi22}.

\begin{figure}[]
\centering
\includegraphics[width=0.47\columnwidth]{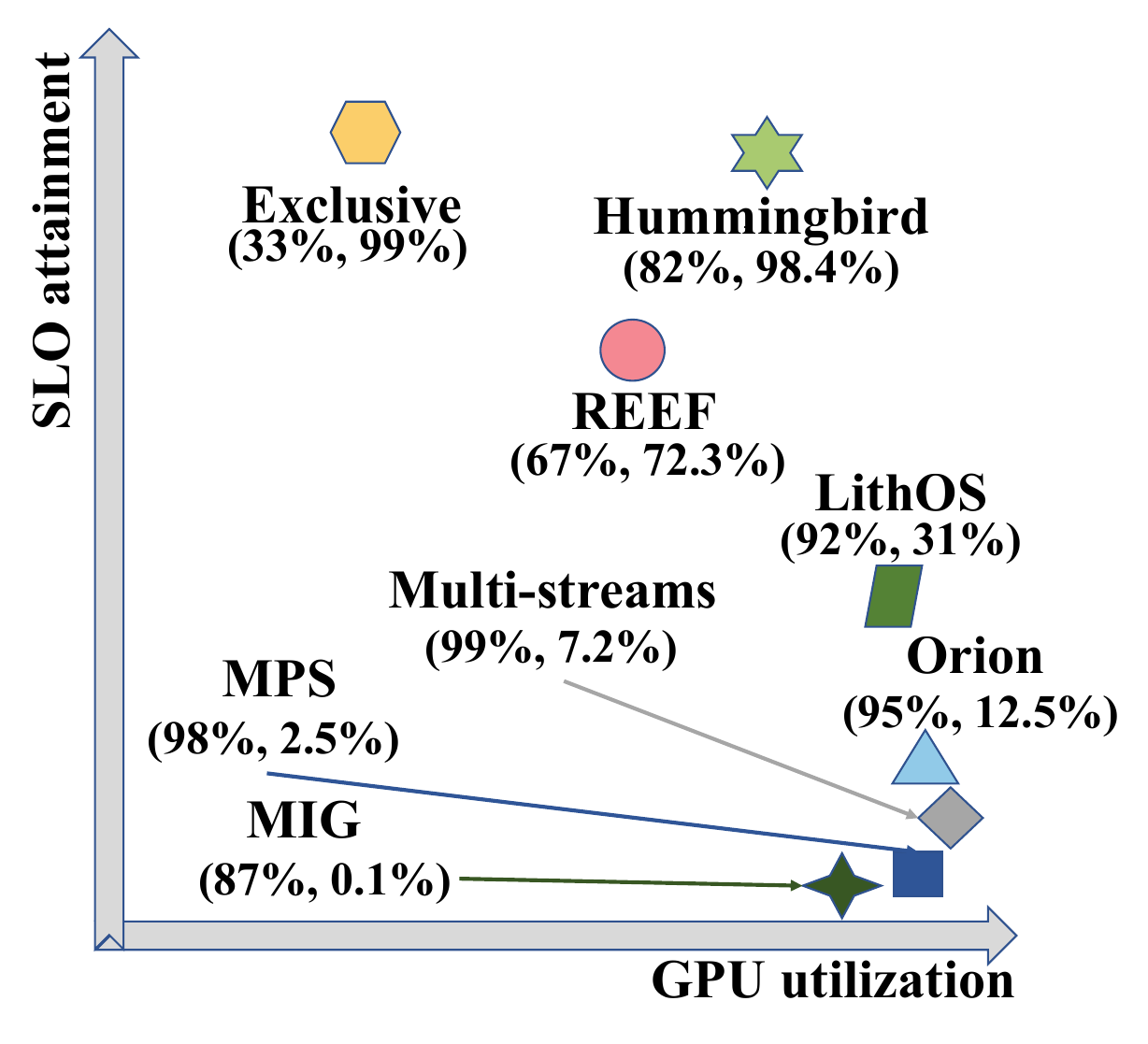}
   \caption{Overview of GPU-sharing techniques. The X-axis represents GPU utilization, while the Y-axis (SLO attainment) represents the proportion of high-priority requests that meet the SLO. \textbf{The SLO is defined as the 99\upth percentile (P99) latency for exclusive execution} as in previous works~\cite{Autothrottle, conserve}.}
  \label{fig:related work summary}

\end{figure}

Although a series of GPU-sharing techniques~\cite{lithos, REEF, MIG, multi_stream, orion@eurosys24} have been proposed to improve GPU utilization by collocating different tasks on the same GPU, they cannot handle the tension well by either failing to adhere to the SLO goal or under-utilizing the GPUs, as summarized in Figure~\textcolor{DarkGreen}{\ref{fig:related work summary}}.

\MyPara{State-of-the-art.}
GPU sharing has been extensively studied in previous works, and there are two types of GPU sharing: spatial sharing~\cite{MIG,multi_stream, Zico@ATC21, orion@eurosys24} and temporal sharing~\cite{REEF, TGS@NSDI23, tally}. 

Spatial sharing, such as the CUDA GPU Streams (referred to as \emph{multi-streams})~\cite{multi_stream}, allows different tasks to execute simultaneously on the same GPU by launching multiple streams. The CUDA kernels within a stream must execute in sequence, while the kernels of different streams can be scheduled to run on different Stream Multiprocessors (SM) concurrently~\cite{stream-multiprocessors}, improving the GPU utilization through improving the intra/inter-SMs level parallelism.
Although the programmers can prioritize the latency-critical tasks by assigning higher priority to the corresponding streams, the spatial sharing techniques can hardly provide SLO guarantees due to severe interference, which is affected by multiple factors and is difficult to control. The inherent reason is that the closed-source GPU, \ie, NVIDIA, does not provide the ability of \emph{fine-grained} task scheduling and resource isolation to developers, leaving a series of hardware resources, such as the GPU L2 cache, HBM bandwidth, and PCIe bandwidth, uncontrolled by users during the spatial sharing (See \S\textcolor{DarkGreen}{\ref{sec:motivation}}). 

Although the interference can be eliminated through statically partitioning the GPU resources, \ie, Multi-Instance GPU (MIG)~\cite{MIG}, it can not dynamically re-partition the resources according to applications' phase-changing behaviors, which results in either low GPU utilization or a series of SLO violations due to the lack of computability. Methods like LithOS \cite{lithos} and Green Context \cite{green_ctx} provides
TPC-level compute control \cite{TPC_mask}, but the other resources, \eg, HBM bandwidth and L2 cache interference, still remain uncontrolled.

\begin{figure}
    \centering
    \includegraphics[width=0.21\textwidth]{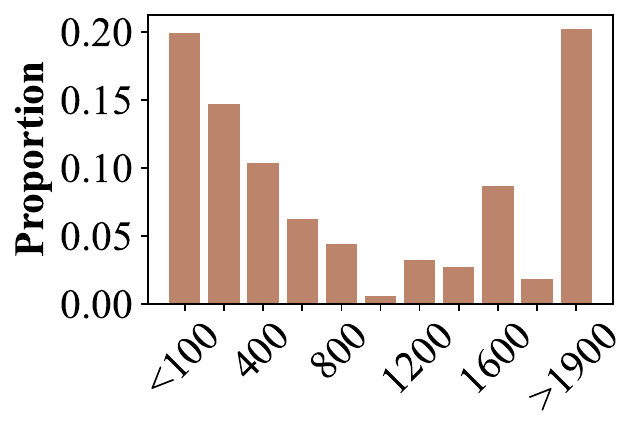}
    \includegraphics[width=0.23\textwidth]{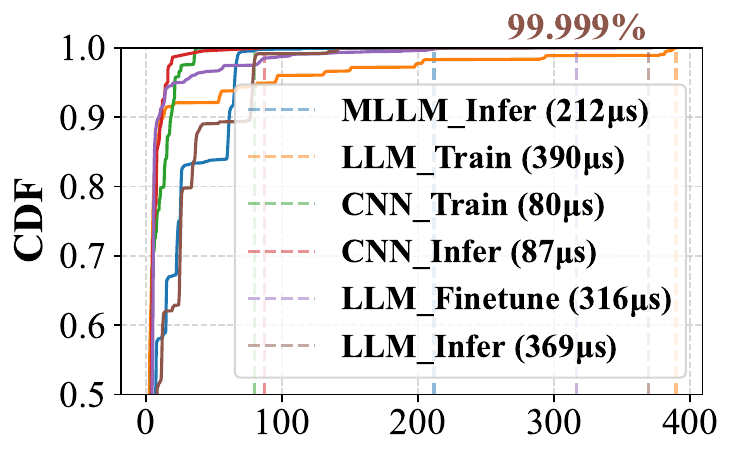}
    \caption{(a) The distribution of kernel execution time (\us); (b) The CDF of the thread block execution time (\us).}
    \label{fig:insight}
\end{figure}

Temporal sharing, such as REEF~\cite{REEF}, is more latency-oriented and provide better SLO attainment by allowing tasks to occupy the GPU exclusively. However, current temporal sharing techniques still show gaps in GPU utilization and SLO attainment. First, since the request rates for high-priority tasks are highly variable \cite{Workload, Antman}, low-priority tasks are frequently preempted and rescheduled to execute, leading to nontrivial synchronization and relaunch overhead (\S\ref{sec:motivation} for more details). 
Second, since the NVIDIA GPU does not support \emph{proactive preemption}~\cite{lack_preemption}, the high-priority tasks have to wait for the completion of any running low-priority kernels, which could last many milliseconds and violate SLO.
For Llama-8B inference on an NVIDIA A100 GPU, kernel execution times span from 5\us for simple elementwise operations to 7.49ms for complex matrix multiplications, \eg, GEMM kernels, resulting in unpredictable preemption time as summarized in Figure~\textcolor{DarkGreen}{\ref{fig:insight}(a)}.

For a series of online services, \eg, ChatGPT, guaranteeing SLO is particularly critical for user satisfaction and revenue, so the industry tends to reserve redundant GPU resources to cope with the burst requests, leaving the GPU cluster underutilized \cite{gpu-workload-analysis@atc19, mlaas@nsdi22, Onlinesole1, mooncake}. Hence, a challenge arises in the GPU data center\textemdash \emph{how to maximize GPU utilization at the premise of ensuring SLO goals?}

\MyPara{Major insights.}
Based on the fact that NVIDIA GPUs are dominating AI applications, from inference to training, the design of GPU scheduling must consider hardware limitations and align application behaviors with hardware characteristics. The core design principle of SLO-oriented scheduling is to ensure that high-priority tasks execute with strict performance isolation, while low-priority tasks opportunistically harvest idle time slices. Critically, the low-priority tasks must release the GPUs in a timely manner, \ie, quitting execution at the \us-scale.
After analyzing dozens of popular AI workloads, we find that this goal is achievable due to the following reasons: 


First, although the execution time of a kernel can span milliseconds, the duration of a single thread block is typically on the scale of microseconds, as each block processes only a small subset of the work to maximize parallelism. It is possible to regulate and minimize the duration for which a low-priority task occupies the GPU to just microseconds by launching tasks at the block level rather than the kernel level. By adjusting the number of blocks launched (referred to as \emph{split-kernel}), the GPU time consumed by low-priority tasks can be finely controlled, creating a series of \emph{preemption points} that allow the scheduler to pause the low-priority tasks and reschedule the high-priority tasks at the granularity of \us-scale, ensuring the SLO of high-priority tasks. As shown in Figure~\textcolor{DarkGreen}{\ref{fig:insight}(b)}, on the NVIDIA A100, the execution times of 99.999\% blocks are within 390\us for a wide spectrum of AI workloads, from CNN~\cite{Resnet, MobileNet}, LLM~\cite{Llama, GPT} to MLLM~\cite{llava, shareGPT4V}, from inference, fine-tuning to training. It is practical to limit the kernel execution time within 400\us by adjusting the number of blocks.

\htc{
Second, the finely controlled low-priority tasks provide more opportunities to improve the GPU utilization by filling the bubbles (\ie, idle GPU time slices).
The bubbles can be categorized into two types. On the one hand, the fluctuation of requests leads to a series of large bubbles, \ie, ranging from seconds to minutes, accounting for up to 23.6\% GPU time in the real-world GPT serving trace \cite{burstgpt}.
On the other hand, there are a huge number of underutilized small bubbles, \ie, at the scale of hundreds of microseconds, during the execution of high-priority tasks. The small bubbles can be divided into three types: memory operations and synchronizations, inter-GPU communication, and CPU-side bound.}

\begin{figure}[]
    \centering
    \includegraphics[width=0.47\textwidth]{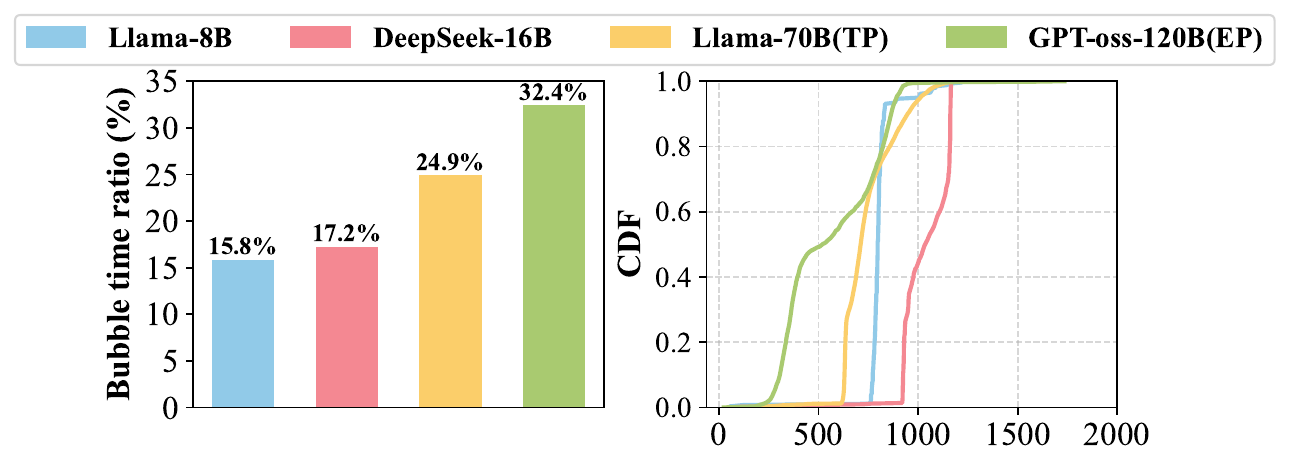}
    \caption{(a) The proportion of small bubbles when processing a request; (b) The distribution of small bubble time (\us).}
    \label{fig:bubble}
\end{figure}

\htc{
For instance, Figure~\textcolor{DarkGreen}{\ref{fig:bubble}(a)} reveals that over 15\% of the GPU time is consumed by such bubbles during Llama-8B~\cite{Llama} and DeepSeek-16B~\cite{deepseek} inference. These bubbles typically originate from the iteration-level device-host memory operations and synchronizations, such as transferring generated tokens for streaming responses and updating batch metadata for continuous batching. While prior works~\cite{zerooverlap} attempt to mitigate these bubbles via overlap techniques, they often incur side effects, such as increased latency or throughput degradation. Furthermore, the proportion of small bubbles can be significantly enlarged by 1.8$\times$ in the distributed setup. Consequently, models like Llama-70B under tensor parallelism and GPT-oss-120B \cite{gptoss120b} under expert parallelism have more than 24\% GPU time consumed by small bubbles due to frequent inter-GPU data transfer and synchronization. To demonstrate the pervasive nature of small bubbles in production environments, we conducted a comprehensive study across 6 models and 6 frameworks, spanning LLM inference and training. Our results confirm that microsecond-scale bubbles are ubiquitous in real-world workloads, as detailed in Supplementary Materials~(\S\textcolor{DarkGreen}{A.1}).}

\MyPara{\name.}
This paper introduces an SLO-oriented GPU scheduling system, \emph{\name}, that allows high-priority tasks to perform preemption on closed-source GPUs, \ie, NVIDIA, at microsecond-scale, while maximizing the GPU utilization through harvesting the idle GPU time slices.
\name contains three components: First, a kernel splitter analyzes the characteristics of low-priority kernels and the capabilities of the underlying hardware to determine the optimal splitting size for each kernel, generating detailed splitting logs to guide the runtime scheduler. Second, a runtime scheduler leverages these logs to split kernels into smaller split-kernels to enable \us-scale preemption. It also dynamically detects idle GPU bubbles, adaptively consolidates split-kernels during large bubbles, and employs a kernel-tick scheduling policy to enhance system throughput. Finally, a memory management module incorporates NVLink \cite{Nvlink} to enable hierarchical memory offloading in GPU sharing.

The design of \name is to tackle two challenges:

First, \emph{how to find the optimal size when splitting a kernel}? 

A smaller kernel contributes a lower preemption delay and creates more opportunities to fill the bubbles with varied sizes, but may underutilize the GPU and introduce more kernel launching overhead. 
\name adjusts the kernel size by controlling the number of blocks within it during kernel launching. The shortest execution time of the split kernel (referred to as the \emph{optimal split-kernel}) is achieved when the number of kernel thread blocks is aligned with the GPU computability\textemdash just fill up the SMs (or saturate the GPU HBM bandwidth). Fewer blocks only reduce GPU utilization due to the idle Tensor/CUDA cores~\cite{a100} but maintain the same kernel execution time.  The optimal splitting size is affected by both the kernel itself and a series of hardware specifications. 
Hence, the splitter will profile the low-priority tasks and automatically calculate the optimal splitting size by considering both SM and bandwidth limitations. The profiling has trivial overhead since most DNNs are iterative~\cite{clockwork@OSDI20,orca}. To make kernel splitting automatic and general, we implement a PTX-based kernel transformer and integrate \name to the real-world frameworks like PyTorch (\S\textcolor{DarkGreen}{\ref{subsec:kernel-spliter}}).

Second, \emph{how to maximize the GPU utilization while ensuring SLO}?

\name improves the GPU utilization by frequently harvesting the idle GPU time slices and avoiding frequent synchronization overhead of low-priority tasks~\cite{REEF}. Request processing latency typically ranges from hundreds of milliseconds to several seconds~\cite{clockwork@OSDI20, Distserve}, several orders of magnitude larger than the average split-kernel execution time of 77\us. Given that detected bubbles span from hundreds of microseconds to milliseconds (Figure~\textcolor{DarkGreen}{\ref{fig:bubble}(b)}), they provide ample capacity to accommodate these split-kernels. Consequently, the scheduler can harvest these bubbles without incurring perceptible latency penalties. Additionally, synchronization is necessary to constrain the GPU device queue length and ensure that preemption delays remain within the execution time of a split-kernel. To mitigate this overhead, \name{} employs a kernel-tick scheduling policy and dynamically consolidates split-kernels when the scheduler detects large bubbles. (\S\textcolor{DarkGreen}{\ref{subsec:scheduler}}).

\MyPara{Results.}
We have evaluated \name with two high-priority tasks and four low-priority tasks, covering a wide spectrum of applications, ranging from CNN to LLM, from inference to training, on three types of GPUs, from mid-end L40s~\cite{l40s} to high-end A100~\cite{a100} and H100~\cite{h100}. In addition, we have evaluated how \name performs under memory-intensive cases and distributed settings by using up to sixteen GPUs of two NVIDIA DGX A100 640GB servers~\cite{a100dgx}.
The evaluation results demonstrate that \name significantly outperforms the state-of-the-art spatial-sharing (\ie, Orion~\cite{orion@eurosys24} and LithOS~\cite{lithos}) and temporal-sharing (\ie, REEF~\cite{REEF}) solutions, achieving 9.7$\times$ and 3.5$\times$ higher SLO attainment, respectively. When compared to executing exclusively, the SLO attainment of the high-priority task only drops by less than 1\%. Meanwhile, \name significantly improves the throughput of low-priority tasks by 2.4$\times$ through improving GPU utilization compared to REEF. The comprehensive experiments prove that \name can be readily used in today’s complex GPU clusters with excellent generality and performance.

\section{Background and Observations}

\subsection{Characterizing Kernel Time in DNNs}
\label{subsec:kernel-diversity}

\MyPara{Diversity.}
The kernel execution time in DNNs exhibits significant heterogeneity. As shown in Figure~\textcolor{DarkGreen}{\ref{fig:insight}(a)}, kernel execution times are distributed across a wide range, reflecting the varying complexity of operations in different DNN tasks. According to our statistics, the minimal kernel is only several microseconds, but the maximal kernel can reach up to tens of milliseconds. This heterogeneity is mainly caused by the varying degrees of parallelism and the intrinsic computational complexity. For example, matrix multiplication is computationally intensive and highly parallel, utilizing a large number of threads across GPU SMs. In contrast, the vector addition operation is lightweight with lower parallelism, as it processes fewer elements simultaneously.

Despite the inherent heterogeneity in GPU workloads, the execution time of thread blocks tends to be consistently short. As shown in Figure~\textcolor{DarkGreen}{\ref{fig:insight}(b)}, our analysis of diverse models, including CNNs~\cite{Resnet, MobileNet}, LLMs~\cite{Llama, GPT}, and MLLMs~\cite{llava, shareGPT4V}, across inference, fine-tuning, and training, reveals that 99.999\% of thread blocks complete execution within 400\us. This observation suggests that \emph{it is feasible to regulate kernel execution times within the threshold of \us-scale by adjusting the number of blocks}. While this analysis demonstrates the general trend, blocks exceeding 400\us may still occur, as they depend on the programmer. To address this issue, some compiler-based approaches \cite{tensor_compiler} have been proposed to automatically partition and reorder kernels, thereby reducing block execution times. These approaches are orthogonal and could serve as complementary solutions.



\MyPara{Predictability.} 
Although the kernel execution time may vary as the input changes, the thread block execution time is highly predictable due to its stability and the iterative nature of DNN workloads. The DNN kernel block is the basic programming abstraction of CUDA and is primarily designed for deterministic linear algebra computations such as matrix multiplication. They typically lack conditional branches or variable loops. This regular computing pattern ensures consistent execution time, which is also confirmed in previous research~\cite{REEF, orion@eurosys24, spatial_2}. Taking the chatbot~\cite{chatbot1} as an example, changes in user prompt length translate to varying parallelism at the kernel level. Common solutions include changing the number of kernels or the number of blocks within a kernel. We have observed similar solutions in popular inference engines, \eg, \emph{llama.cpp}~\cite{llama.cpp} and \emph{SGLang}~\cite{sglang}, which do not change the execution time of thread blocks. Furthermore, DNN workloads are inherently iterative~\cite{clockwork@OSDI20, Rammer@OSDI20}, repeatedly executing the same kernel blocks across training or inference steps. Profiling an iteration can accurately predict future execution characteristics.


\subsection{Hardware Limitations of GPU}
\label{subsec:hardware-limitations}

Most closed-source GPUs, in particular NVIDIA GPUs, do not allow users to preempt kernels after submission \cite{lack_preemption, orion@eurosys24}. The GPU hardware scheduler dispatches thread blocks from kernels in each work queue based on stream priority, which remains uncontrollable by users. These limitations present significant challenges for designing GPU-sharing techniques. The basic GPU sharing mechanisms supported by NVIDIA either struggle to achieve robust resource isolation, \eg, MPS~\cite{MPS}, or fail to support dynamic resource allocation, \eg, MIG~\cite{MIG}, which underpins why spatial sharing cannot guarantee low latency for high-priority tasks. 
For example, spatial-sharing approaches such as Orion \cite{orion@eurosys24} aim to co-schedule kernels with complementary resource requirements, but their coarse-grained kernel-level scheduling falls short compared to block/warp-level GPU hardware scheduling, making them inadequate for mitigating interference when two heavy applications run concurrently. Although some works \cite{xsched, REEF_TOCS} propose an undocumented \codeIn{ioctl} to realize kernel interruption, it can not be restored and is only valid in specific architectures.
\section{Motivation}
\label{sec:motivation}

In this section, we quantitatively demonstrate why the spatial-sharing solution fails to meet the SLO and the significant overhead of existing temporal-sharing solutions. 

\MyPara{Setup.} 
We perform Llama-8B \cite{Llama} inference with real-world trace, BurstGPT \cite{burstgpt}, as the high-priority task and conversation summarization using the ShareGPT dataset \cite{shareGPT, chatgpt} as the low-priority task, a typical offline inference scenario \cite{conserve}, executed with Mistral-7B \cite{mistral}. 
Both models are implemented in \textit{llama.cpp} \cite{llama.cpp} with 8-bit quantization. The high-priority task uses the serving mode of \textit{llama.cpp} with default settings, and the low-priority task is set as offline inference with a batch size of 32 to optimize the throughput. We use the state-of-the-art solutions, \ie, Orion~\cite{orion@eurosys24} and REEF~\cite{REEF}, as the representations of spatial-sharing and temporal-sharing solutions, respectively. The parameters are configured the same as in their original papers~\cite{orion@eurosys24,REEF}. We evaluate the performance of Llama-8B inference when executing exclusively on the A100 as a baseline. The SLO is defined as the 99\upth percentile latency under exclusive high-priority task execution by following previous research~\cite{Autothrottle, conserve}.

\begin{figure}
    \centering
    \includegraphics[width=0.8\linewidth]{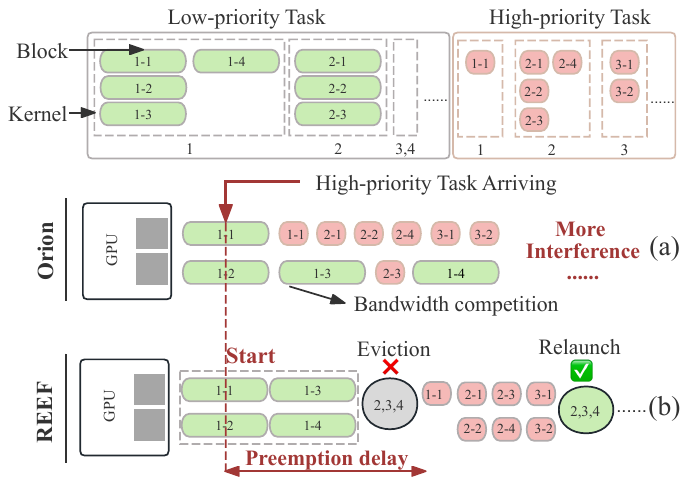}
    \caption{An illustration of GPU task scheduling across two co-run schemes: (a) Orion, (b) REEF.}
    \label{fig:motivation-case-study}
\end{figure}

 \begin{figure}[]
        \centering
    \includegraphics[width=0.20\textwidth]{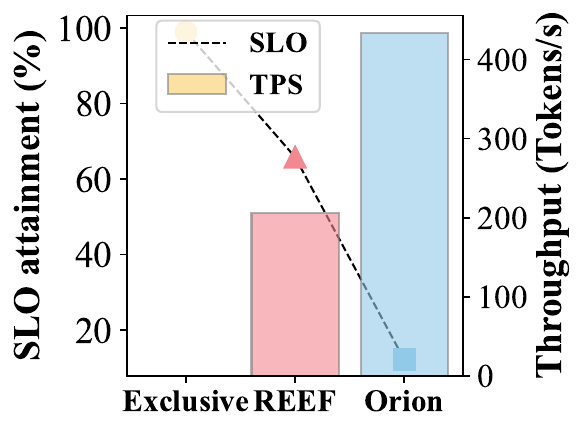}
         \includegraphics[width=0.20\textwidth]{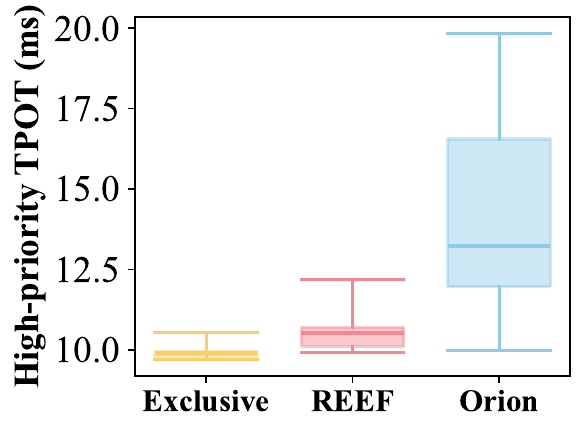}
         \caption{(a) SLO attainment of high-priority task (gray lines) and throughput of low-priority tasks (bars-TPS); (b) Boxplot of the time per output token (TPOT) of high-priority tasks.}
         \label{fig:motivation-data}
 \end{figure}

\MyPara{Interference within the spatial sharing.}
As shown in Figure~\textcolor{DarkGreen}{\ref{fig:motivation-data}(a)}, Orion's SLO attainment of the high-priority task is 5.4$\times$ and 8.1$\times$ lower than REEF and the exclusive mode. Orion can only mitigate the interference to a certain extent by selectively collocating kernels from high- and low-priority tasks with opposite resource requirements, but it cannot eliminate the interference. We have observed severe memory bandwidth contention between the collocating tasks. 

As shown in Figure~\textcolor{DarkGreen}{\ref{fig:motivation-data}(b)}, when running on Orion, the time variation of generating a token in Llama-8B inference is enlarged by 4.2$\times$ and 10.9$\times$ compared to running on REEF and executing exclusively, respectively. The root causes are as follows\textemdash first, the profiling and classification of the \emph{compute-intensive} and \emph{memory-intensive} kernels are too coarse-grained. For example, Orion marks a kernel as memory-bounded only when its average memory bandwidth utilization exceeds 60\%. As a result, the selected compute-intensive kernels can still saturate the GPU global memory bandwidth when collocating with the Llama-8B kernels during the decoding phase, which is hugely memory-intensive; second, Orion is implemented on the \textit{multi-streams}~\cite{multi_stream}, whose kernel scheduling policy is opaque to users. There is no guarantee that the collocated compute-intensive and memory-intensive kernels shall start and exit at the same time. For example, as shown in Figure~\textcolor{DarkGreen}{\ref{fig:motivation-case-study}(a)}, the stranded memory-intensive blocks (block\codeIn{<1-3>}) of the low-priority task may co-run with the subsequent memory-intensive blocks (block\codeIn{<2-1>}and\codeIn{<2-2>}) of the high-priority task, which causes bandwidth interference. Besides, the SM occupancy delays the execution of other high-priority blocks. The stranded block problem poses a severe challenge to spatial sharing, a weakness that is exacerbated by modern AI tasks, which include thousands of kernels in one execution.

\MyPara{Preemption delay.}
REEF achieves 5.4$\times$ better SLO attainment due to its temporal sharing strategy, which significantly alleviates the interference compared with Orion. However, REEF still exhibits SLO violations (underperforms the exclusive mode by 1.5$\times$, Figure~\textcolor{DarkGreen}{\ref{fig:motivation-data}(a)}) due to the preemption delay. There are two sources for the delay. First, when the high-priority task arrives, it has to wait for the completion of the low-priority kernel that is executing, \eg, from block\codeIn{<1-1>} to block\codeIn{<1-4>}. Such a delay depends on the execution time of low-priority kernels, which is diverse and affected by a series of factors in real-world DNNs. Some kernels, such as matrix multiplications, can take several milliseconds to execute, which is much larger than the bubbles of high-priority tasks, as in \S~\textcolor{DarkGreen}{\ref{subsec:kernel-diversity}}, resulting in unpredictable preemption delay. Second, the kernels buffered in the device queue need to check the \emph{preemption flag} and voluntarily quit their execution (\ie, the eviction~\cite{effisha@ppopp17, REEF} procedures in Figure~\textcolor{DarkGreen}{\ref{fig:motivation-case-study}(b)}). The cost of eviction is strongly related to the number of buffered kernels. Hence, the existing solutions usually restrict the capacity of the device queue, \ie, four kernels in REEF, but this will cause more kernel launching and synchronization overhead.

\MyPara{Throttled throughput of low-priority tasks.}
Although REEF proactively restores the low-priority tasks when the GPU is idle, the throughput of low-priority tasks is significantly throttled, which underperforms Orion by 2.1$\times$, as shown in Figure~\textcolor{DarkGreen}{\ref{fig:motivation-data}(a)}. The fundamental reasons are the inability to utilize small bubbles and high-frequency synchronization overhead. In our experiments, 23.4\% of low-priority task computations are wasted due to kernel eviction and relaunch. After preemption, the evicted kernels must be relaunched, resulting in additional overhead due to the frequent CPU-to-GPU instruction transfers. In extreme cases, \eg, low-priority kernel \codeIn{<4>} in Figure~\textcolor{DarkGreen}{\ref{fig:motivation-case-study}(b)}, need to be relaunched up to three times due to frequent preemption. Besides, to mitigate excessive kernel eviction and relaunch, REEF caps the GPU’s device queue length at four, but it introduces non-trivial synchronization overhead, which accounts for 6.8\% of GPU cycles.




\MyPara{Key Takeaway.}
Spatial-sharing introduces significant interference to high-priority tasks, usually failing to meet SLO. Although temporal sharing can provide better SLO attainment, its effects are also significantly influenced by hardware limitations and workload characteristics, leaving opportunities for further improvement. In addition, the good latency of the temporal sharing is achieved at the expense of a throttled throughput of low-priority tasks. 

\section{Design}
\label{sec:design}

\subsection{Overview}
\label{subsec:overview}

In this section, we will discuss how \name can address these challenges by splitting and regulating the heterogeneous kernels of low-priority tasks to ensure \us-scale preemption and maximize the GPU utilization by filling bubbles and kernel-tick scheduling.

\begin{figure}[]
    \centering
    \includegraphics[width=0.65\linewidth]{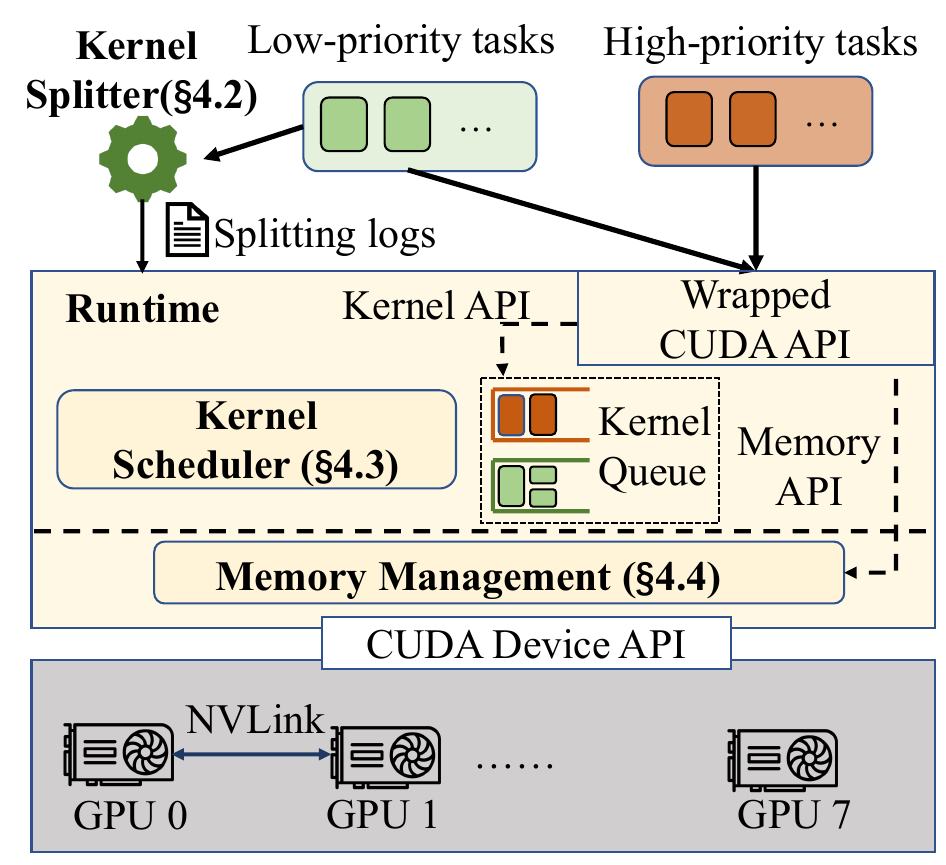}
    \caption{Design overview of \name{}.}
    \label{fig:overview}
\end{figure}

\begin{figure}[]
    \centering
    \includegraphics[width=0.75\linewidth]{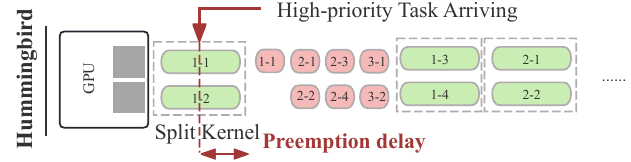}
    \caption{An illustration of GPU task scheduling of \name. The tasks here are the same as Figure~\ref{fig:motivation-case-study}.}
    \label{fig:our scheduling}
     \vspace{-10pt}
\end{figure}

As shown in Figure~\textcolor{DarkGreen}{\ref{fig:overview}}, \name contains three components: (1) A \emph{kernel splitter} that is used to analyze the kernel execution time of low-priority tasks and calculate the optimal kernel splitting size based on hardware features. Then, when high-priority tasks arrive, the preemption latency (\ie, the remaining execution time of split kernels) is limited to \us-scale, as shown in Figure~\ref{fig:our scheduling} (\S\textcolor{DarkGreen}{\ref{subsec:kernel-spliter}}); 
(2) a runtime scheduler that dynamically splits and consolidates the kernel according to the size of GPU bubbles to balance the preemption latency and GPU utilization (\S\textcolor{DarkGreen}{\ref{subsec:scheduler}});
and (3) an NVLink-extended memory management system that supports hierarchical memory offloading (\S\textcolor{DarkGreen}{\ref{subsec:unified-memory}}).


\subsection{Kernel Splitter}\label{subsec:kernel-spliter}

In this section, we discuss how \name finds the \emph{optimal split-kernel size} by analyzing the GPU hardware features, such as the computability and memory bandwidth. 

\MyPara{Optimal split-kernel size.}
As aforementioned, a smaller kernel size (of low-priority tasks) helps reduce the preemption latency and provides more opportunities to fill the small GPU bubbles of high-priority tasks. However, a kernel size that is too small will underutilize GPU resources. Hence, the key insight in identifying the optimal splitting size lies in understanding the relationship between kernel execution time and GPU utilization. The optimal (shortest) execution time of a split-kernel is achieved when the number of kernel threads is aligned with the GPU computability\textemdash just fill up the SMs (or saturate the GPU global memory bandwidth, which will be discussed later). At the same time, fewer kernel threads will leave part of the SMs idle, but will not reduce the execution time of the kernel. \htc{We employ a two-step analysis method to calculate the optimal split-kernel size.} 


First, we calculate the maximum number of thread blocks that a split-kernel should contain by considering the SMs, which is calculated by:
\begin{equation}
    N_{\textit{block}} = N_{\textit{SM}} \cdot o \cdot \frac{ SM\_MAX\_THREADS}{THREADS\_PER\_BLOCK}
    \label{eqa:block_number}
\end{equation}

$N_{\textit{SM}}$ is the number of SMs, and $SM\_MAX\_THREADS$ is the number of threads within a SM. They are both parameters specific to the hardware.  $THREADS\_PER\_BLOCK$ is specified by the program developer. $o$ refers to kernel occupancy, which is depicted by the number of threads an SM can handle concurrently, depending on kernel properties (\eg, shared memory and register usage).

Second, starting from the number of blocks, \ie, $N\_\textit{block}$, calculated in the previous step, we gradually reduce the block count while observing the kernel's execution time. If reducing the number of blocks results in a shorter execution time, the kernel is memory-bound, as fewer blocks reduce contention for memory bandwidth due to fewer concurrent memory access instructions. The reduction continues until the execution time stabilizes, signaling that the kernel has reached its shortest execution time, and then we can calculate the optimal splitting size.
The results are recorded in splitting logs, which guide the runtime scheduler in efficiently splitting, consolidating, and managing kernel execution.

\begin{figure}[]
    \centering
    \includegraphics[width=0.8\linewidth]{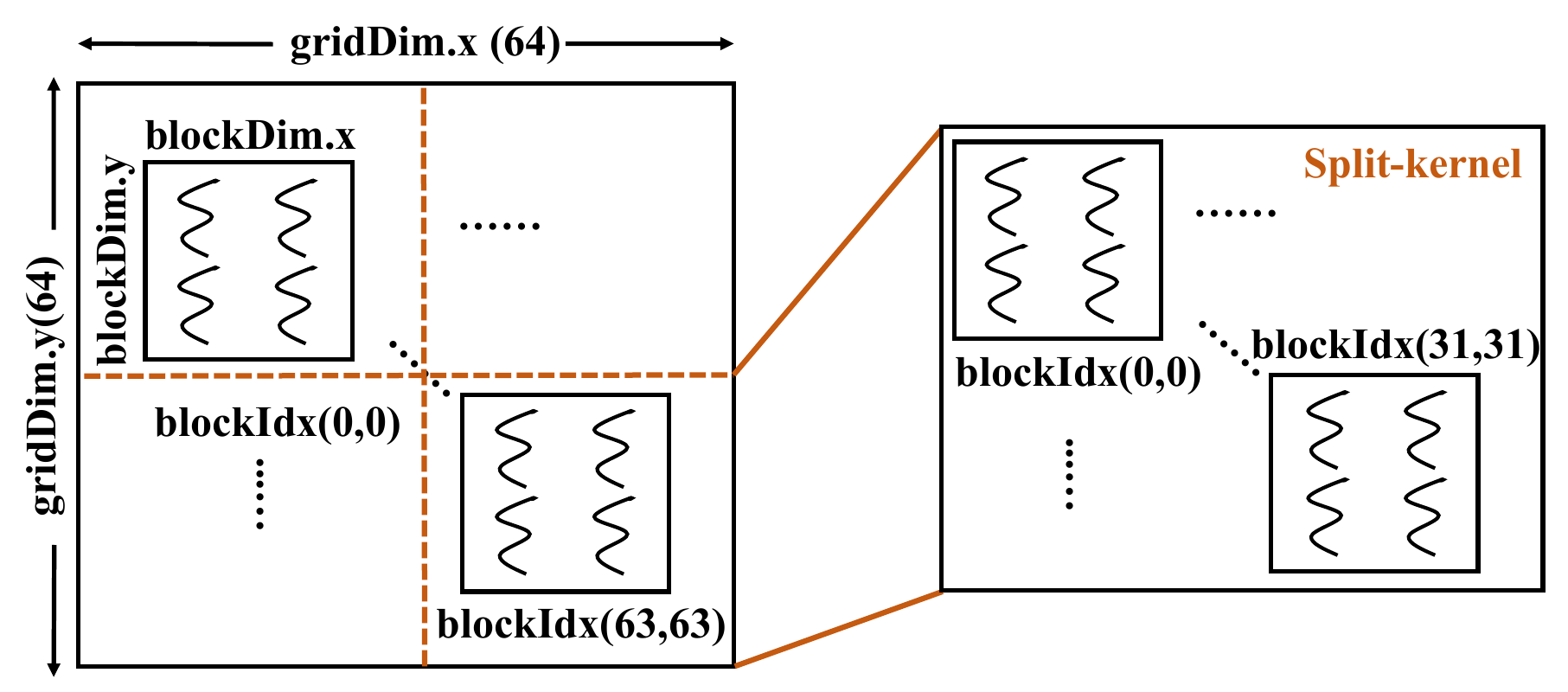}
    \caption{An example of kernel splitting.}
    \label{fig:cuda-program-example}
\end{figure}

\MyPara{PTX kernel transformation.} 
\htc{ In this section, we detail the mechanism of enabling kernel splitting achieved via automatic transformation of kernel device code (PTX assembly~\cite{ptx})}. \htc{ GPU computations are organized as kernels, each composed of multiple thread blocks managed within a grid. These blocks execute independently and are uniquely identified by the built-in \codeIn{blockIdx} variable. Figure~\textcolor{DarkGreen}{\ref{fig:cuda-program-example}} illustrates a representative kernel from Llama-8B inference, comprising 4096 thread blocks arranged in a (64, 64) 2D grid, where \codeIn{blockIdx.x} and \codeIn{blockIdx.y} range from 0 to 63.}

\begin{lstlisting}[language=c++, 
caption={Example of kernel transformation.}, 
float=h, 
moreemph={int, size_t, struct, void, __global__}, 
commentstyle=\color{ForestGreen}, % 设置注释为绿色,
emphstyle=\bfseries \color{blue},
morekeywords={(blockIdx.x + offset[0]), (blockIdx.y + offset[1]), offset[3], cudaLaunchKernel},
keywordstyle=\color{purple},
label={lst:source code transformation}]
.visible .entry mulmat(
    .param  .u64    mulmat_param_0, 
    .param  .u64 mulmat_param_1, ..., 
    (*@\colorbox{gray!15}{.param    .u32    mulmat\_param\_offset\_x, ..y, ..z}@*)){
    .reg .b32 	%r<6>; // Declare registers
    (*@\colorbox{gray!15}{ld.param.u32  \%r1,   [mulmat\_param\_offset\_x];}@*)
    (*@\colorbox{gray!15}{ld.param.u32  \%r2,   [mulmat\_param\_offset\_y];}@*)
    ... // load other parameters
    mov.u32     %r3, %ctaid.x;
    (*@\colorbox{gray!15}{add.s32  \%r3, \%r3, \%r1;}@*)
    ... // left kernel body
    }
\end{lstlisting}


\htc{ Splitting a monolithic kernel into smaller sub-kernels necessitates realigning the \codeIn{blockIdx} to maintain correctness. For instance, splitting the aforementioned kernel into four sub-kernels of size (32, 32) reduces the local index range to [0, 31], breaking the original mapping. To address this, we employ a PTX injection technique, conceptually similar to eBPF~\cite{eBPF, NEUTRINO}, to instrument the kernel code. Specifically, we modify the PTX to accept additional offset parameters and inject arithmetic instructions to shift the native \codeIn{blockIdx} (represented as \codeIn{ctaid} in PTX) by these offsets, thereby preserving the original addressing semantics. Listing~\textcolor{DarkGreen}{\ref{lst:source code transformation}} demonstrates this transformation on a \textit{mulmat} operator. Further implementation details are provided in \S\textcolor{DarkGreen}{\ref{sec:implementation}}.}

\htc{ We choose PTX-level code transformation mainly for generality. From the perspective of compilation, the complicated GPU ecosystem can roughly be divided into two branches: Ahead-of-Time (AOT) compiled operator libraries of hand-written codes (\ie, CUDA), and Just-in-Time (JIT) compiled domain-specific languages (DSLs) such as Triton \cite{triton}. These two branches unify at the parallel assembly layer (PTX in NVIDIA) \cite{NEUTRINO}, so we choose this level to ensure generality. We validate its generality by integrating \name into real systems such as PyTorch \cite{pytorch} and successfully transforming complex kernels in CUTLASS \cite{cutlass} and Triton \cite{triton}.}

\subsection{Runtime Scheduler}\label{subsec:scheduler}
\name's runtime scheduler manages kernel execution, such as detecting bubbles, dynamically splitting and consolidating the kernels to balance
the preemption latency and GPU utilization, as shown in Algorithm~\textcolor{DarkGreen}{\ref{alg:scheduling}}.

\setlength{\textfloatsep}{0.1cm}
\begin{algorithm}[]
\small
\KwIn{
High(Low)-priority kernel queue $Q_{\text{hp}}$ / $Q_{\text{lp}}$; 
}
\DontPrintSemicolon
\SetKwFunction{Main}{\textsc{KERNELSCHEDULING}}
    \SetKwProg{Fn}{Function}{:}{}
    \Fn{\Main{$Q_{\text{lp}}$, $Q_{\text{hp}}$}}{
    \While{$True$}{
        \If{$!Q_{\text{hp}}.is\_empty()$}{ \label{alg:hp_start}
            $P_{\text{flag}} \leftarrow True$  \textit{\textcolor{DarkGreen}{// Preemption flag}} \;
            $Launchkernel (Q_{\textit{hp}})$\;
        }   \label{alg:hp_end}
        \Else{ \label{alg:lp_start}
            $Bubble_{\text{flag}} \leftarrow \textsc{DETECTBUBBLES}()$\; \label{alg:bubble_detect}
            \If{$Bubble_{\text{flag}} == False$} {  
            $Continue$ \textit{\textcolor{DarkGreen}{// Do not find bubbles}}\; 
            }
            \If{$Is\_large(Bubble_{\text{flag}})) == True$} { \label{alg:Consolidate_start}
            $\textsc{CONSOLIDATE} (Q_{\text{lp}})$\;} \label{alg:Consolidate_end}
            $GPU\_sync()$ \textit{\textcolor{DarkGreen}{// Wait for high-priority finish}}\; \label{alg:sync}
            $P_{\text{flag}} \leftarrow False$\;
            \textit{\textcolor{DarkGreen}{/* Asynchronous thread to launch low-priority kernels and exit when $P_{\text{flag}}$ is set to True */}}\;
            $Call$ $\textsc{KERNEL\_TICK} (Q_{\text{lp}}, P_{\text{flag}}) $ $thread$ \;  \label{alg:launch}
        } \label{alg:lp_end}
    } 
}
\caption{Kernel scheduling logic (simplified)} 
\label{alg:scheduling}
\end{algorithm}
\setlength{\textfloatsep}{0.1cm}

\MyPara{4.3.1  \textit{High-priority kernel scheduling}}
\label{section:Online scheduling}
\
\newline
When high-priority kernels arrive, the scheduler immediately halts the launch of any new low-priority kernel and launches high-priority kernels (Lines~\textcolor{DarkGreen}{\ref{alg:hp_start}-\ref{alg:hp_end}}).  In our evaluation, \name achieves an average preemption delay of 139\us, resulting in less than 1\% slowdown to request processing latency, which is a negligible overhead. (\S\textcolor{DarkGreen}{\ref{subsec:preemption delay}}).



\MyPara{4.3.2  Low-priority kernel scheduling}
\label{sec:lp-scheduling}
\
\newline
The scheduling of low-priority kernels must address two key challenges: determining \emph{when to launch the low-priority kernels} and \emph{how to consolidate the split-kernels}. 
Blindly launching the low-priority kernels can either result in excessive unnecessary preemption, significantly harming the SLO of high-priority tasks, or miss opportunities to utilize idle GPU bubbles, leading to GPU underutilization.


\MyPara{Kernel splitting.}
Upon receiving a kernel launch function call (\eg, \texttt{cuLaunchKernel}) from low-priority tasks, the scheduler queries the logs from the kernel splitter and splits the kernel into small \emph{split-kernels} according to the optimal splitting size. The scheduler calculates the grid size and \codeIn{blockIdx} offset for each split-kernel and enqueues them for scheduling.

\MyPara{Bubbles detection.}
Low-priority kernels are scheduled only when the scheduler detects bubbles on the GPU. This detection occurs when no high-priority kernels are pending in the runtime kernel queue (Lines \textcolor{DarkGreen}{\ref{alg:lp_start}-\ref{alg:bubble_detect}}). The bubbles can be divided into two types: (1) small bubbles, and (2) large bubbles. 


\htc{ Small bubbles primarily stem from cross-device data transfers and synchronization (\eg, CPU-GPU or inter-GPUs). While typically lasting several hundred microseconds or a few milliseconds, these bubbles occur orders of magnitude more frequently than large bubbles, manifesting as idle or underutilized intervals on the GPU timeline. Exploiting these bubbles necessitates a mechanism capable of real-time detection. However, this task is complicated by the absence of online, microsecond-level GPU monitors. To address it, we propose a novel hint-based bubble detection mechanism.}

\htc{
We find that the small bubbles can be identified through host-side hints. Within specific frameworks and applications, each bubble type is characterized by deterministic API patterns. For instance, the bubble during each iteration of LLM inference, as aforementioned in \S\textcolor{DarkGreen}{1}, always begins with a CUDA API \codeIn{cudaMemcpyAsync} followed by \codeIn{cudaStreamSynchronize} pattern. Similarly, communication bubbles in pipeline parallelism are typically marked by NCCL~\cite{NCCL} APIs such as \codeIn{ncclSend/Recv}\footnote{While some APIs are completed via kernels, these kernels will not be interfered with by low-priority kernels. Details in Supplementary Materials.}. \name inserts marker events (start/end) before the host launches via lightweight \codeIn{cudaEvent}. When the scheduler detects the start event, it confirms the start of small bubbles. Through such a host-side hint-assisted bubble detection mechanism, \name can detect small bubbles effectively. We conducted a comprehensive study across 6 models
and 6 frameworks, spanning LLM inference and training, to demonstrate the pervasive nature of small bubbles in production environments (Supplementary Materials (\S\textcolor{DarkGreen}{A.1})). Furthermore, to generalize this approach, we developed an automated tool to identify the patterns of small bubbles by parsing Nsight Systems~\cite{nsight-system} traces (Supplementary Materials (\S\textcolor{DarkGreen}{A.2})).
}



Large bubbles, ranging from tens of milliseconds to seconds, are common in serving workloads that are caused by request fluctuation or network latency. Our scheduler periodically scans the GPU device queue and identifies a large bubble when no high-priority kernel appears within a time threshold. We set the threshold just above the observed duration of small bubbles for different high-priority tasks and frameworks. Upon detecting a large bubble, the scheduler consolidates kernels by restoring the original grid size and resetting offsets (Lines \textcolor{DarkGreen}{\ref{alg:Consolidate_start}-\ref{alg:Consolidate_end}}). Although kernel consolidation may increase preemption latency when a high-priority request arrives unexpectedly, large bubbles are far less frequent than small ones. Our experiments confirm that the impact on overall preemption overhead is less than 0.7\%, which is negligible (\S\textcolor{DarkGreen}{\ref{subsec:drill-down}}). To further mitigate it, we adopt a prediction-based consolidation policy. Using a request-interval predictor previously studied \cite{request_prediction}, the scheduler forecasts the next interval to guide kernel consolidation to ensure that the execution time fits within the interval.




\MyPara{kernel-tick scheduling policy.}
After detecting a bubble, the scheduler synchronizes with the GPU to wait for current high-priority kernels to finish. It then sets $P_{\text{flag}}$ to $False$ to indicate the GPU is now executing low-priority kernels, and calls an asynchronous thread that applies the kernel-tick scheduling policy to launch low-priority kernels. This thread stops when $P_{\text{flag}}$ to $True$, which indicates new high-priority kernels are enqueued or bubbles end (Lines \textcolor{DarkGreen}{\ref{alg:sync}}-\textcolor{DarkGreen}{\ref{alg:lp_end}}).

This policy guarantees that the preemption latency is bounded by a single split-kernel execution time by limiting the number of kernels on the GPU device queue to at most one. However, naively synchronizing after every kernel introduces significant synchronization overhead because the synchronization API incurs approximately 5\us overhead, followed by an additional 6–7\us delay due to kernel launch latency. To avoid that cost, \name leverages the predictable execution time of kernels to act as a "tick" scheduler. Instead of synchronizing after each kernel, the scheduler calculates a launch interval based on the kernel execution time minus launch overhead. Using this interval, the scheduler launches the next kernel precisely as the current one nears completion. This approach forms a streamlined CPU-GPU pipeline that minimizes synchronization frequency and reduces overhead. 

\htc{
With these optimizations, \name incurs only a 1.3\% performance slowdown compared with the vanilla implementation without kernel splitting or synchronization, as demonstrated in our evaluation (\S\textcolor{DarkGreen}{\ref{subsec:drill-down}}). Moreover, \name naturally generalizes to an arbitrary number of low-priority tasks by serving them in a round-robin fashion.}

\subsection{Memory Management}
\label{subsec:unified-memory}

\begin{figure}
    \centering
    \includegraphics[width=0.65\linewidth]{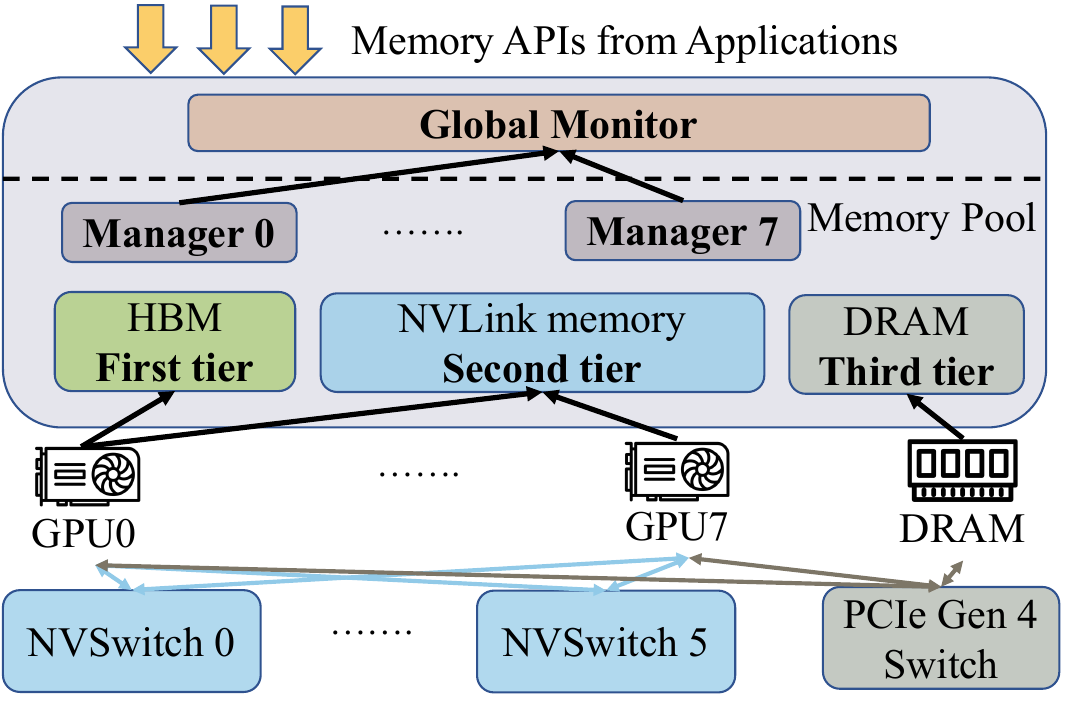}
    \caption{NVLink-extended Unified Memory.  }
    \label{fig:memory manage overview}
\end{figure}

\htc{
As model sizes scale, memory capacity becomes a critical bottleneck for task co-location, a challenge largely overlooked by prior GPU sharing schemes that focus primarily on compute sharing. Our design principle mandates strict isolation: high-priority tasks must retain unhindered access to the full GPU memory capacity, while low-priority tasks opportunistically utilize the residual memory.}

\htc{
Existing techniques, such as TGS \cite{TGS@NSDI23}, which are built upon CUDA Unified Memory \cite{unified_memory} and other DRAM-based offloading mechanisms \cite{pipeswitch, pie, pre_gated_MoE, moe_lightning}, suffer from limited PCIe bandwidth and prolonged memory access latency. GMLake \cite{GMLake} and vAttention \cite{vAttention} construct a dynamic memory management with CUDA virtual memory management (VMM) APIs \cite{VMM} to reduce memory fragmentation. Still, they are not designed for GPU sharing scenarios and lack support for memory offloading.}

\htc{
Previous work, hierarchical unified virtual memory (HUVM) \cite{HUVM}, leveraged the high bandwidth of NVLink to introduce a hierarchical unified memory system, comprising local HBM, NVLink-connected HBM, and DRAM, which enables efficient memory offloading at the GPU driver level. Inspired by it, \name implements an extension of the unified memory management that transparently offloads low-priority task data into idle NVLink-connected memory at the page level, as depicted in Figure~\textcolor{DarkGreen}{\ref{fig:memory manage overview}}. \name applies two optimizations to adapt to GPU-sharing scenarios, which are priority isolation and interference awareness. }

\htc{
First, \name uses placement preferences in CUDA Driver APIs to prioritize the memory allocation of high-priority tasks. When the GPU memory is full, \name only allows for evicting the pages of low-priority tasks. It guarantees that the memory of high-priority tasks will not be swapped, and the GPU memory capacity is the same as the original GPU memory in the view of high-priority tasks.}

\htc{
Second, HUVM adopts a round-robin page eviction policy to maximize the available GPU-to-GPU bandwidth via parallel fetching. However, NVLink shares HBM bandwidth with local kernel execution, and an improper eviction policy can cause severe interference to local high-priority tasks. To mitigate it, \name incorporates a global monitor that measures the real-time bandwidth via a \textbf{ping‑like} method. It first establishes a baseline transmission latency by sending a fixed‑size packet between two GPUs under interference-free conditions and then conducts periodic latency tests during execution. A significant increase over the baseline indicates bandwidth conflict, so the monitor prioritizes swapping to GPUs with lower contention. If NVLink‑connected memory is exhausted or lack of NVLink hardware support, it falls back to offloading pages to DRAM. More details are shown in Supplementary Material (\S\textcolor{DarkGreen}{B}).}

\section{Implementation}
\label{sec:implementation}

\htc{
\name is implemented on NVIDIA GPUs, comprising approximately 8000 lines of C++/CUDA code. \name requires no hardware-specific instructions and supports any generation of GPU architectures.}

\MyPara{CUDA API hook.}
\htc{
\name ensures generality and transparency across diverse ML ecosystems by intercepting low-level CUDA Driver APIs without requiring application modifications, similar to previous work \cite{TGS@NSDI23, lithos}. It redirects APIs like kernel launches (\eg, \codeIn{cuLaunchKernel}) and memory allocations (\eg, \codeIn{cuMemAlloc}) to custom wrappers, thereby enabling fine-grained kernel scheduling and memory management. }

\MyPara{Kernel profiler.}
\htc{
To profile the kernel execution time and select the optimal kernel-splitting size, \name develops a lightweight online profiler to collect kernel information. The profiler uses \codeIn{cudaEvent} to record kernel execution times and corresponding CUDA APIs to acquire hardware features and kernel occupancy. The profiling only leads to trivial overhead because most DNNs are iterative, which takes about several seconds for a task.}

\MyPara{Kernel transformation.}
\htc{
Building upon the probe engine from NEUTRINO~\cite{NEUTRINO}, we implement PTX kernel transformation at runtime. In detail, GPU code, in the ELF [25] or FatBinary [56] format, requires an explicit load via specific APIs (\eg, \codeIn{cuModuleLoad} and \codeIn{cuModuleGetFunction}). We hook these APIs and the probe engine \codeIn{objdumps} the dumped GPU binary to extract PTX and use the kernel name to match and prune the many-kernel raw assemblies into a single-kernel assembly while keeping global definitions and device functions. To enable splitting, we modify the kernel parameter list to accept additional offset parameters, loading them via inserted \codeIn{ld.param} instructions. We then realign the thread block indices by injecting \codeIn{add} instructions to shift the native \codeIn{blockIdx}. Auxiliary tasks, such as register declaration and command line parsing, are handled by NEUTRINO's infrastructure. After probing, the probe engine converts the probed assemblies into machine code via assemblers such as ptxas \cite{ptxas}. The kernel transformation overhead is negligible because every kernel only needs to be injected once.}

\htc{A limitation is that we can not acquire the PTX of closed-source libraries such as cuBLAS \cite{cublas} and cuDNN \cite{cudnn}. To mitigate it, we alternate most of
them with the popular open-sourced library CUTLASS \cite{cutlass}, which has the equivalent performance provided by NVIDIA. LithOS \cite{lithos} introduces a kernel-splitting technique via reverse engineering, but its approach incurs substantially higher overhead than ours due to the large number of early-return short threads. Once LithOS becomes open-source, our framework can incorporate it as a complementary method for closed-source libraries.}

\MyPara{Generality of kernel splitting.}
\htc{
There are a few corner cases of kernel splitting that need special attention. To extend kernel splitting for CUDA graphs \cite{cuda_graph}, \name can intercept graph creation APIs and split graphs into subgraphs, ensuring correct execution ordering, similar to previous work \cite{lithos}. As long as we keep the subgraphs' execution no longer than 400\us, we can still provide a preemption latency guarantee. Some special kernels involving cross-block synchronization (\ie, \codeIn{grid\_group::sync()}) or persistent kernels, \name disables kernel splitting. These kernels can also be solved via source-code level refactoring (\eg, breaking cross-block sync into kernel boundaries).
}
\begin{table*}[]
\centering
\begin{tabular}{c|c|c|c|c|c}
\hline
Model                & Type                                                                                                                                       & Batch Size                                                           & SM Active                                                            & Bandwidth Utilization                                                     & MEM Usage                                                    \\ \hline
\rowcolor[HTML]{FFFC9E} 
{\color[HTML]{000000} Llama-8B (\textbf{LMA}) \cite{Llama}} & {\color[HTML]{000000} LLM Inference}                                       & {\color[HTML]{000000} Serving}                                       & {\color[HTML]{000000} 55.8\%}                                        & {\color[HTML]{000000} 38.7\%}                                        & {\color[HTML]{000000} 12.5\%}                                       \\ \hline

\rowcolor[HTML]{FFFC9E} 
Yi-34B (\textbf{Yi}) \cite{Yi}                       & LLM Inference                                                                                    & Serving                                                                   & 58.3\%                                                               & 53.0\%                                                               & 44.3\%                                                              \\ \hline
\rowcolor[HTML]{C0C0C0} 
Mistral-7B (\textbf{MIST})\cite{mistral}                      & LLM Inference                                                                                        & 32                                                                   & 63.4\%                                                              & 46.5\%                                                               & 12.1\%                                                              \\ \hline
\rowcolor[HTML]{C0C0C0} 
{\color[HTML]{000000} DeepseekMoE-16B (\textbf{DS}) \cite{deepseek}}                          & {\color[HTML]{000000} LLM Inference}                                               & {\color[HTML]{000000} 32}                                       & {\color[HTML]{000000} 75.1\%}                                             & {\color[HTML]{000000} 55.0\%}                                             & {\color[HTML]{000000} 22.5\%}                                             \\ \hline
\rowcolor[HTML]{C0C0C0} 
ResNet101 (\textbf{RN}) \cite{Resnet}                        & CNN Training                                                                     & 64                                                                   & 88.8\%                                                               & 16.4\%                                                               & 12.7\%                                                              \\ \hline
\rowcolor[HTML]{C0C0C0} 
GPT-2 (124M)(\textbf{GPT2}) \cite{GPT}                        & LLM Training                                                                       & 16                                                                   & 91.2\%                                                               & 18.4\%                                                               & 12.4\%                                                              \\ \hline

\end{tabular}
\caption{Evaluation applications. SM active ratio and bandwidth utilization indicate the average SM activity and bandwidth usage during a request (inference) or an epoch (training). Memory usage represents the task's required ratio to total GPU memory. Yellow rows denote high-priority tasks, and grey rows denote low-priority tasks.}
\label{tab:applications}
\end{table*}

\section{Evaluation}
In this section, we first evaluate on the single GPU, which is the default setup of baselines~\cite{orion@eurosys24, REEF} (\S\textcolor{DarkGreen}{\ref{subsec:memory-intensive}}). Then, we evaluated \name under memory-intensive scenarios (\S\textcolor{DarkGreen}{\ref{subsec:memory-intensive}}). Further evaluations, \eg, distributed settings and generality on different types of GPUs, are also evaluated (\S\textcolor{DarkGreen}{\ref{subsec:drill-down}}).



\subsection{Single GPU Performance}~\label{subsec:single-gpu}
\MyPara{Testbed.}
Our single-GPU experiments are conducted on a server equipped with eight A100 (80 GB HBM, SXM4) GPUs, two Xeon(R) 8358P CPUs (total 64 cores), and 1 TB of host memory. The server ran Ubuntu 22.04 and CUDA 12.6. To reduce latency jitter, we turned off dynamic voltage and frequency scaling (DVFS) of GPUs \cite{dvfs1, dvfs2}.

\MyPara{Applications.}
We evaluate six representative workloads spanning from inference to training, as summarized in Table \textcolor{DarkGreen}{\ref{tab:applications}}. High‑priority (\textbf{hp} in short) inference tasks comprise Llama‑8B (\textbf{LMA}) and Yi‑34B (\textbf{Yi}), while low‑priority (\textbf{lp} in short) inference tasks include Mistral‑7B (\textbf{MIST}) and DeepSeekMoE‑16B (\textbf{DS}). We also include two training tasks: ResNet‑101 (\textbf{RN}) and GPT‑2 (\textbf{GPT2}). Llama‑8B and Mistral‑7B are relatively small models with per‑token latencies around 10 ms. DeepSeekMoE‑16B is a recently popular MoE, which dynamically selects experts and thus exhibits greater variability in computational demand. Yi‑34B is a larger, less latency‑sensitive model. Both ResNet-101 and GPT-2 are compute-intensive, where GPT-2 has some long-running kernels that take tens of milliseconds to complete. 

We implement inference tasks using \textit{llama.cpp} \cite{llama.cpp} with 8-bit quantization, and PyTorch \cite{pytorch} for training tasks. We use the serving mode of \textit{llama.cpp} with the real-world trace from BurstGPT \cite{burstgpt} as the workload for high-priority tasks and batched inference or training for low-priority tasks.

\MyPara{Metrics.}
The evaluation compares the SLO attainment of high-priority tasks and the throughput of low-priority tasks. For high-priority tasks, the 99\textsuperscript{th} percentile TTFT and TPOT measured during exclusive execution serve as the SLOs, following the previous work \cite{Autothrottle, conserve}. \emph{SLO attainment} refers to the proportion of requests meeting SLO. GPU utilization refers to the SM active ratio reported by Nsight Systems~\cite{nsight-system}.

\MyPara{Baselines.}
\htc{Our comparative analysis includes REEF~\cite{REEF} as the representative for temporal sharing, and Orion~\cite{orion@eurosys24} and LithOS~\cite{lithos} for spatial sharing. Due to LithOS's closed-source nature, we reconstructed its system by implementing its three key components: dynamic TPC mapping (using \codeIn{libsmctrl}~\cite{TPC_mask}), kernel atomization on open-source kernels, and TPC stealing mechanism. All hyperparameters were set identical to those reported in the LithOS paper. Regarding Orion, we follow its design by assigning high-priority tasks to the highest-priority CUDA stream and low-priority ones to the default stream. REEF is deployed with a device queue capacity of 4, consistent with its default configuration. }

\begin{figure}[]
    \centering
    \includegraphics[width=0.48\textwidth]{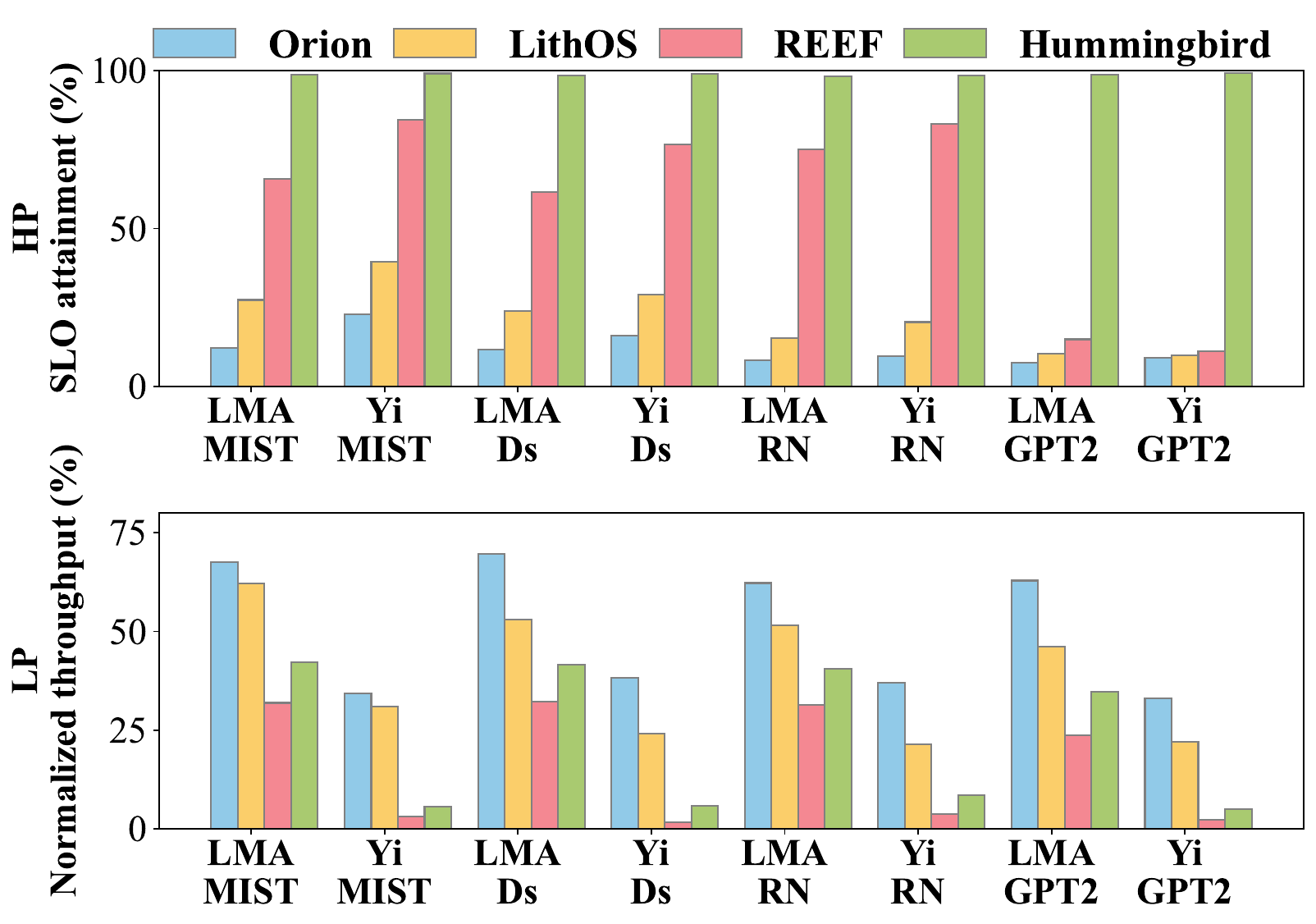}
    \caption{(a) SLO attainment of high-priority tasks; (b) Throughput of low-priority tasks. The results are normalized to executing low-priority tasks exclusively.}
    \label{fig:SLO_eval}
\end{figure}

\MyPara{SLO attainment of high-priority tasks.}
Figure~\textcolor{DarkGreen}{\ref{fig:SLO_eval}(a)} illustrates the SLO attainment of high-priority tasks. Orion, employing spatial sharing, performs poorly across all cases, with no case exceeding 22.8\% SLO attainment. Especially when the low-priority tasks are training tasks, the SLO attainment does not exceed 10\% due to the more intense competition for computing resources. The root cause lies in its kernel-level scheduling, which is too coarse-grained and cannot align with the fine-grained scheduling mechanisms of GPU hardware, such as block or warp-level scheduling (\S~\textcolor{DarkGreen}{\ref{sec:motivation}}). \htc{Although LithOS shows 1.8$\times$ better SLO attainment than Orion due to its TPC mapping and kernel atomization to allow for more fine-grained compute control and kernel scheduling, it still has significant limitations. Specifically, it fails to ensure that the running low-priority tasks yield resources promptly. Furthermore, severe interference persists due to contention for shared memory resources, specifically HBM bandwidth and L2 cache. This combination of untimely resource release and unmanaged memory interference constitutes the root cause preventing spatial sharing from guaranteeing the SLOs. }

In contrast, REEF delivers significantly better SLO attainment due to its temporal-sharing mechanism. On average, it achieves 2.7$\times$ higher SLO attainment of high-priority tasks than Orion and LithOS, but its performance remains suboptimal for certain models, especially for Llama-8B. This model has short per-token iteration times, making it latency-sensitive and easily affected by preemption delays. The issue is further exacerbated when low-priority tasks involve GPT training, where the SLO attainment drops below 14.9\%. This is due to some long-running kernels in the GPT training task, which significantly increase REEF’s preemption latency.


\name achieves nearly 99\% SLO attainment in all scenarios thanks to microsecond‑scale preemption.
On average, it outperforms LithOS and REEF by $5.6\times$ and $3.0\times$ in SLO attainment, respectively. Regardless of whether low‑priority workloads are batched inference or long‑running training kernels, \name caps maximum preemption delay at 400\us through kernel splitting and controlled scheduling, incurring under 1\% slowdown on high‑priority tasks. As a result, \name’s microsecond-scale preemption remains effective, demonstrating its practicality in guaranteeing the SLO of high-priority tasks under real‑world deployments.

\MyPara{Throughput of low-priority tasks.}
Figure~\textcolor{DarkGreen}{\ref{fig:SLO_eval}(b)} shows the normalized throughput of low-priority tasks. While Orion and LithOS achieve the higher throughput due to their spatial sharing nature to enhance GPU utilization, it severely compromises the SLO attainment of high-priority tasks.

Compared with REEF, \name achieves 1.9$\times$ higher throughput of low-priority tasks on average. This improvement is attributed to \name's ability to utilize small bubbles effectively, whereas REEF struggles with frequent kernel evictions and significant relaunch overhead when it fills the small bubbles. Additionally, \name mitigates the synchronization overhead through the kernel-tick scheduling policy. When the high-priority task is Yi-34B, \name achieves higher throughput speedup ($2.5\times$) compared to REEF. This is because Yi-34B has longer request processing times, resulting in a higher proportion of small bubbles of the GPU time slices allocated to low-priority tasks. Additionally, we measure the GPU utilization of \name and REEF, which are 82\% and 67\%, respectively, further explaining why \name achieves better throughput.



\subsection{Memory-intensive Cases}~\label{subsec:memory-intensive}
\MyPara{Settings.} 
To evaluate the benefits of NVLink‑extended memory management, we configure Llama‑8B and Yi‑34B inference as high‑priority tasks and Llama‑70B inference as the low‑priority task. These two combinations need around 90 GB and 115 GB, requiring
approximately 13\% and 47\% of the low-priority task’s memory to be swapped out, respectively. We use Orion, LithOS, and REEF with HUVM as baselines. 

\begin{figure}
    \centering
    \includegraphics[width=0.42\textwidth]{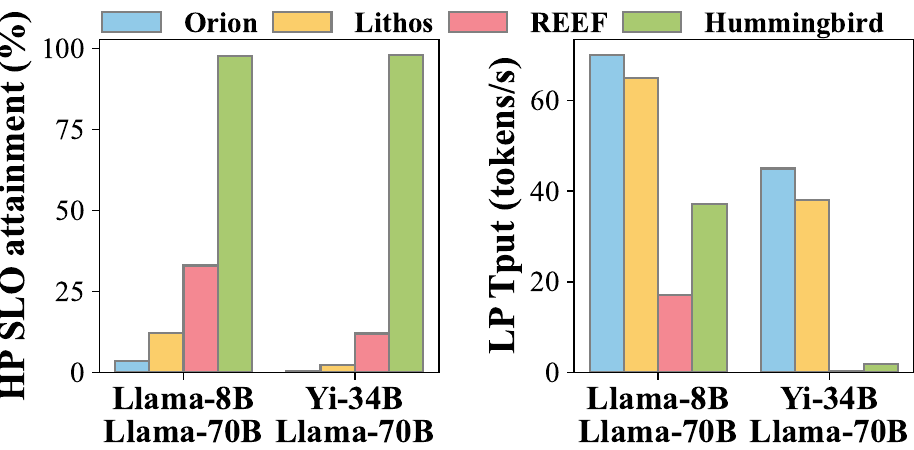}
    \caption{(a) SLO attainment of high-priority tasks; (b) Throughput of low-priority tasks.}
    \label{fig:memory_case}
\end{figure}

\MyPara{Results.} \htc{As shown in Figure~\textcolor{DarkGreen}{\ref{fig:memory_case}(a)}, Orion and LithOS exhibit poor high-priority SLO attainment, peaking at only 12\%. The root cause lies in their spatial sharing mechanism, which triggers intense memory contention, leading to frequent page swapping and severe access latency. Moreover, this increased latency exacerbates the stranded block problem, delaying the release of low-priority resources. Conversely, \name achieves a 5.6$\times$ improvement compared to REEF, maintaining high attainment with negligible degradation (<3\%). This performance gap stems from our optimized memory management, specifically priority isolation and interference-aware eviction. Whereas, REEF further suffers from extended execution of low-priority kernels and high preemption overhead.}


\htc{
Figure~\textcolor{DarkGreen}{\ref{fig:memory_case}(b)} shows the throughput of the low-priority task. \name outperforms REEF by 4.2$\times$. The main benefits come from \name's ability to effectively utilize idle GPU time slices, while REEF's kernel eviction and relaunch overhead are amplified. The main cause is that REEF's kernel eviction mechanism has to check a flag that is stored in HBM, but it may be swapped out due to memory contention. }

\htc{
Results in memory-intensive environments (a critical challenge for GPU sharing in the era of large models) demonstrate that \name can still strongly guarantee the SLO of high-priority tasks while utilizing small bubbles to improve the throughput of low-priority tasks, further proving the effectiveness of \name in cloud GPU clusters. }


\subsection{Performance Drill Down}~\label{subsec:drill-down}
\begin{figure}[]
    \centering
        \includegraphics[width=0.42\textwidth]{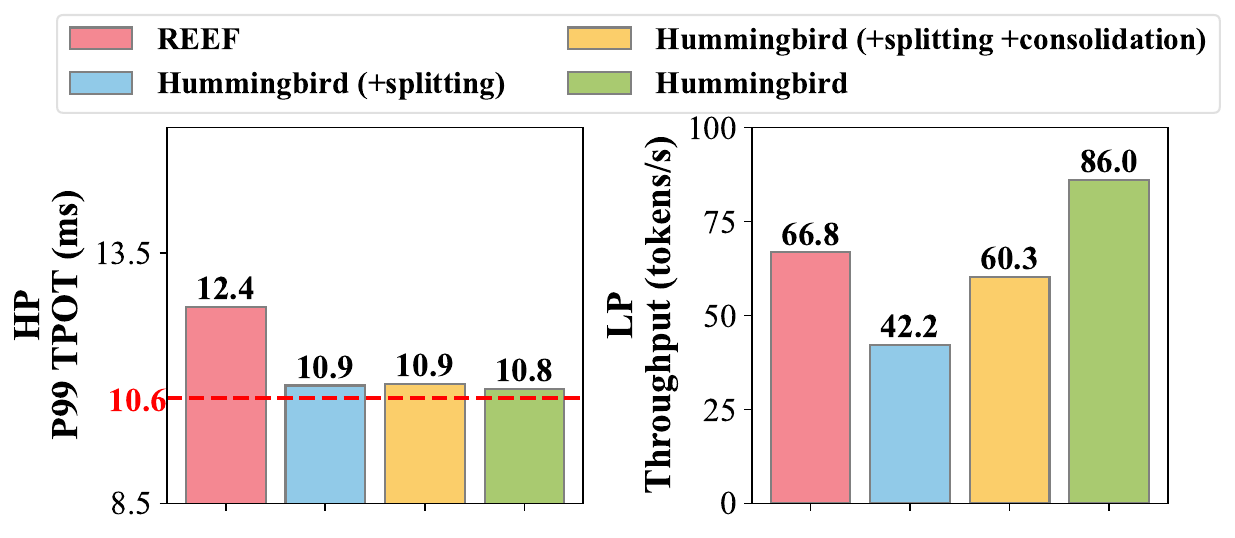}
    \caption{\name's optimizations break down. (a) The 99\upth TPOT of high-priority tasks. The red dashed line represents the 99\upth TPOT of high-priority tasks when running exclusively. (b) The throughput of low-priority tasks.}\label{fig:time-breakdown}
\end{figure}

\MyPara{Breaking down end-to-end speedup.}
We re-evaluated the Llama-8B and Mistral-7B case to demonstrate the effectiveness of our three key optimizations: kernel splitting, kernel consolidation, and scheduling policy, as shown in Figure~\textcolor{DarkGreen}{\ref{fig:time-breakdown}}.

For high‑priority tasks, REEF’s 99\upth TPOT rises to 12.5ms, whereas \name’s kernel splitting reduces it to 10.9ms, coming close to the 10.6ms achieved when high‑priority work runs exclusively. In addition, kernel consolidation has only a minimal impact on the SLO of high-priority tasks, which is less than 0.7\% on overall preemption latency. This is because large bubbles primarily occur between requests, whose preemption delay is only a very small part compared to the high-frequency preemption caused by small bubbles.

For low‑priority tasks, kernel splitting initially incurs a 37\% throughput slowdown. To recover this loss, \name applies kernel consolidation during large bubbles, which reduces synchronization and launch overhead and boosts throughput by 1.5$\times$. In addition, frequent synchronization still causes up to a 10.7\% slowdown to low-priority tasks, so \name introduces a kernel‑tick scheduling policy to control the kernel launch timing, which reduces synchronization frequency and improves throughput by $1.43\times$. These techniques improve low‑priority throughput by $1.29\times$. When compared with a vanilla implementation without any kernel splitting or synchronization, \name introduces only a 1.3\% slowdown.


\begin{figure}[]
    \centering
    \includegraphics[width=0.2\textwidth]{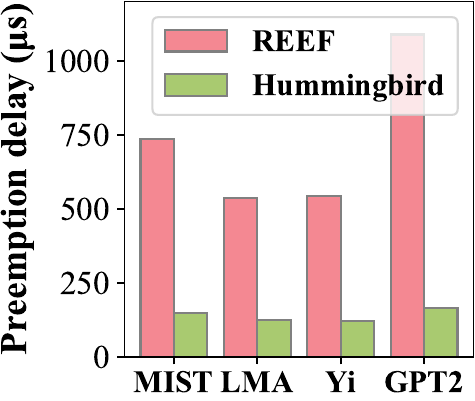}
    \hspace{1.5em}
    \includegraphics[width=0.2\textwidth]{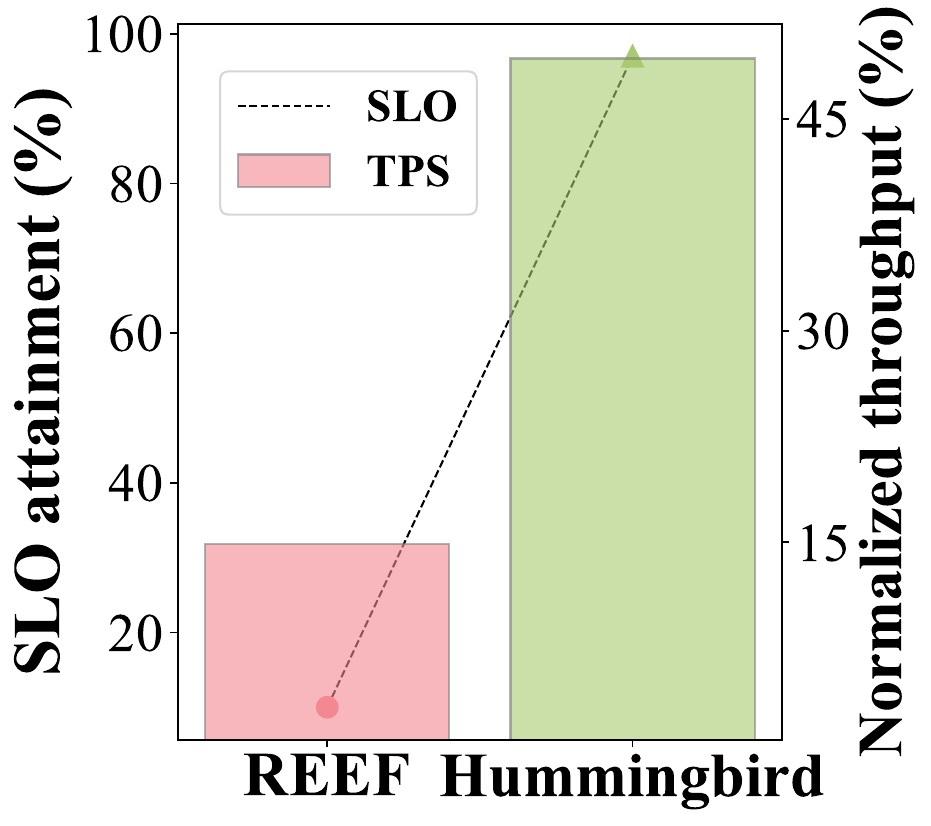}
    \caption{(a) Comparison of average preemption latency; (b) Performance comparison under multi-GPUs scenarios.}
    \label{fig:preemption latency}
\end{figure}

\MyPara{Preemption delay.}\label{subsec:preemption delay}
Figure~\textcolor{DarkGreen}{\ref{fig:preemption latency}(a)} shows the average preemption delay for various workloads. \name reduces latency substantially, achieving between 121\us and 165\us and speedups from 4.3$\times$ to 6.6$\times$ compared to REEF. This improvement arises from \name’s ability to split kernels and limit the kernel execution time below 400\us, while REEF must wait for heterogeneous kernels to complete before preemption, resulting in much higher and unpredictable delays.

\MyPara{\name in Multi-GPUs.}
\name can scale to multi-GPU environments with minimal modifications, because each GPU runs its kernel scheduler that is entirely transparent to the application, where distributed policies are application-level concerns. To evaluate \name's scalability, we use sixteen A100 GPUs on AWS EC2 of two \texttt{p4de.24xlarge} instances~\cite{aws}, each equipped with 8 NVIDIA A100 80GB GPUs. GPUs within the same node communicate via 600 GB/s NVLink 3.0, while inter-node communication relies on 400 Gbps UltraFast Ethernet. We deploy Llama-405B using a hybrid parallelism strategy (\ie, tensor parallelism within nodes and pipeline parallelism across nodes) as high-priority tasks. We use GPT-2 training of distributed data parallelism (DDP) for low-priority tasks. We mainly evaluate \name against REEF, which is extended to support distributed settings in a per-GPU kernel scheduler fashion. Spatial sharing is omitted due to its extremely low SLO attainments.

As shown in Figure~\textcolor{DarkGreen}{\ref{fig:preemption latency}(b)}, \name achieves a $9.7\times$ improvement in SLO attainment compared to REEF. The high degree of tensor parallelism (8 GPUs per pipeline stage) significantly accelerates inference, making the application highly sensitive to preemption latency. REEF's long preemption delay results in poor SLO performance. In contrast, \name effectively addresses these challenges through its \us-scale preemption and maintains low latency and high SLO attainment. As for the throughput of low-priority tasks, \name delivers $3.3\times$ higher throughput than REEF. The primary cause is REEF's inability to effectively harvest small execution bubbles. This limitation is exacerbated in distributed environments, where frequent inter-GPU data transfer and synchronization generate a large number of small bubbles, accounting for up to 30\% of total execution time, consequently leading to suboptimal GPU utilization.

\MyPara{Generalization to other GPUs.}
To show that \name can generalize to other GPUs, we re-evaluate the Llama-8B and Mistral-7B case on L40s and H100. 

\begin{figure}
    \centering
    \includegraphics[width=0.45\textwidth]{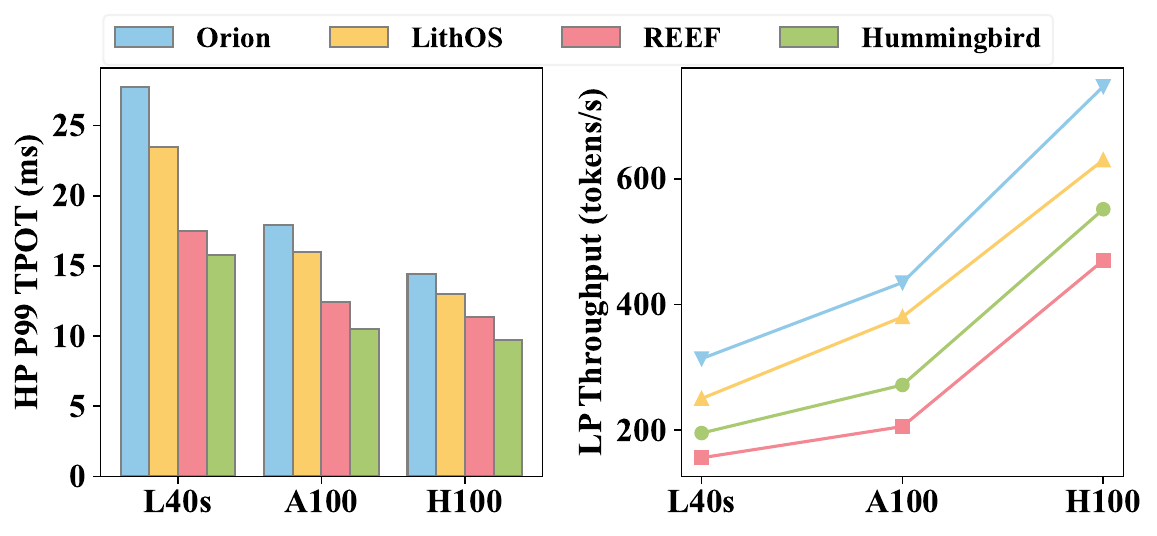}
    \caption{(a) 99\upth TPOT of high-priority task, and (b) throughput of low-priority task across different GPUs.}
    \label{fig:multigpu}
\end{figure}

As shown in Figure~\textcolor{DarkGreen}{\ref{fig:multigpu}}, \name achieves 39.0\%, 31\% and 13.3\% lower 99\upth TPOT compared to Orion and REEF, and achieves similar tail latency to exclusive mode. On the one hand, as the GPU computing power increase, the 99\upth TPOT of all four GPU sharing techniques decreases (lower is better). 
On the other hand, for Orion and LithOS, more sufficient resources mean less competition and improved the 99\upth TPOT. For REEF, the kernel execution time becomes shorter, leading to a smaller preemption delay. 

The throughput of low-priority tasks follows a similar trend. While Orion and LithOS achieves the higher throughput, thet do so at the cost of increased latency for high-priority tasks. In contrast, \name delivers $1.25\times$ higher throughput than REEF without compromising the performance of high-priority tasks. These results demonstrate that \name can generalize across different GPUs, ensuring the SLO of high-priority tasks while maximizing GPU utilization.

\section{Related Works}\label{sec:related-work}

\MyPara{Spatial Sharing.}
Spatial sharing mechanisms \cite{spatial_1, spatial_2, spatial_3, spatial_4, laius@SC19, MIG, MPS, Zico@ATC21, orion@eurosys24, Wavelet@Mlsys21, miso@SC22} allow multiple jobs to utilize distinct GPU regions concurrently. NVIDIA Multi-Instance GPU (MIG) \cite{MIG} supports hardware-level partitioning but lacks flexibility for dynamic resource reclamation during idle periods, with reconfiguration taking hundreds of milliseconds and checkpoint recovery tens of seconds \cite{miso@SC22}. NVIDIA Multi-Process Service (MPS) \cite{MPS} enables concurrent execution but often suffers from interference due to shared access to caches, compute units, and bandwidth.
Methods like Zico \cite{Zico@ATC21} and Tick-Tock \cite{Wavelet@Mlsys21} optimize memory use by coordinating forward and backward passes but fail to support diverse workloads or prioritize tasks. Interference-aware solutions, such as Orion \cite{orion@eurosys24} and BLESS \cite{BLESS}, address contention at the kernel level, but either struggle with resource-intensive applications running simultaneously or overlook the priority of tasks. LithOS \cite{lithos} allows for more fine-grained compute control and kernel scheduling but still fail to solve the bandwidth interference.
Conserve \cite{conserve} enables co-scheduling of high and low priority LLM requests on a single GPU with strict SLO guarantees. However, it supports only a single model and cannot handle mixed-model deployments or heterogeneous task types.

\MyPara{Temporal Sharing.}
Temporal sharing techniques \cite{temporal_1, temporal_2, temporal_3, gandiva@osdi18, Salus@Mlsys20, clockwork@OSDI20, Antman, TGS@NSDI23, REEF, tally} divide GPU time into slices, enabling context switching for better utilization. Approaches like Gandiva \cite{gandiva@osdi18} suspend and resume models, moving states between GPU and host memory, and Antman \cite{Antman} dynamically adjusts memory allocations for efficient colocation. Clockwork \cite{clockwork@OSDI20} precomputes deadlines to achieve predictable latency, and TGS \cite{TGS@NSDI23} provides application-agnostic sharing for containerized workloads, simplifying integration with diverse systems. Gpreempt \cite{Gpreempt} proposes a time-slice based preemption, but lacks support for LLM and application-level observations.REEF \cite{REEF} implements task preemption for commodity GPUs, but due to the closed-source nature of NVIDIA GPUs, it cannot forcibly kill running kernels and instead waits for them to finish (referred to as REEF-N). XSched \cite{xsched} proposes a general preemptive scheduler across XPUs, but its implementation on NVIDIA GPUs relies on kernel eviction and relaunch, which is similar to REEF. These eviction-based approaches work for smaller workloads such as ResNet \cite{Resnet} but struggle with the resource demands, tight latency requirements, and dynamic behavior of large LLM workloads, underscoring the need for more robust, fine-grained temporal sharing mechanisms. Although some works \cite{xsched, REEF_TOCS} propose undocumented kernel interruption methods, they can not restore interrupted kernels and are only valid in the specific architectures (i.e., Volta \cite{volta}), which lack generality.

\section{Conclusion}
This paper presents \name, an SLO-oriented GPU
scheduling system that allows high-priority tasks to perform preemption on closed-source GPUs, \ie, NVIDIA, at microsecond-scale, while maximizing the GPU utilization. 
Our promising results demonstrate that \name{} can be readily used in today’s GPU clusters.

\bibliographystyle{plain}
\bibliography{references}

\clearpage
\appendix
\section*{Supplementary Materials}
\addcontentsline{toc}{section}{Supplementary Materials}
\begin{table*}[]
\small
\centering

\begin{tabular}{c|c|c|c|c}
\hline
Id & Model                & Framework             & Bubble pattern                                                                                                                       & Bubble duration                              \\ \hline
1  & Qwen2.5-7B Inference & vLLM/llama.cpp/SGLang & \begin{tabular}[c]{@{}c@{}}cudaMemcpyAsync+cudaStreamSynchronize\\ for streaming response\end{tabular}                               & 500\us-1000\us \\ \hline
2  & Qwen2.5-7B Inference & vLLM/llama.cpp/SGLang & \begin{tabular}[c]{@{}c@{}}A series of cudaMemcpyAsync\\ for continuous batching\end{tabular}                                        & 700\us-1500\us \\ \hline
3  & GPT2 Training    & DeepSpeed            & \begin{tabular}[c]{@{}c@{}}A large chunk of cudaMemcpyAsync\\ for allreduce gradients\end{tabular}                                   & about 6ms                                    \\ \hline
4  & GPT2 Training    & DeepSpeed/Megatron   & \begin{tabular}[c]{@{}c@{}}cudaMemcpyAsync+cudaDevice(Stream)Synchronize \\ for forward metadata preparation\end{tabular}            & 400\us-600\us  \\ \hline
5  & GPT2 Training    & Megatron              & \begin{tabular}[c]{@{}c@{}}A series of cudaMemcpyAsyc+cudaHostAlloc\\ for checkpoint store every 100 training terations\end{tabular} & 300\us                        \\ \hline
\end{tabular}
\caption{The bubble summarization of the memory operation and synchronization type.}
\label{tab: memory type bubble}
\end{table*}

\begin{table*}[]
\small
\centering
\begin{tabular}{c|c|c|c|c|c}
\hline
Id & Model                                   & Parallelism & Framework                                                           & Bubble pattern                                                                                                                  & Bubble duration                              \\ \hline
6  & \multirow{6}{*}{Llama3-70B Inference}             & TP          & llama.cpp                                                           & \begin{tabular}[c]{@{}c@{}}cudaMemcpy3DPeerAsync+cudaStreamWaitEvent\\ for output aggregation\end{tabular}                      & about 200\us                  \\ \cline{1-1} \cline{3-6} 
7  &                                         & TP          & \begin{tabular}[c]{@{}c@{}}vLLM/SGLang/\\ TensorRT-LLM\end{tabular} & ncclAllReduce for output aggregation                                                                                            & 500\us-1500\us \\ \cline{1-1} \cline{3-6} 
8  &                                         & PP          & \begin{tabular}[c]{@{}c@{}}vLLM/SGLang/\\ TensorRT-LLM\end{tabular} & ncclSendRecv for Inter-stage communication                                                                                      & 150\us-1000\us \\ \hline
9  & \multirow{2}{*}{GPT-oss-120B Inference} & TP+EP       & vLLM/SGLang                                                         & custom cross-device reduce for TP/EP aggregation                                                                                & 230\us-10ms                   \\ \cline{1-1} \cline{3-6} 
10 &                                         & DP+EP       & vLLM                                                                & ncclBroadcast for expert routing                                                                                                & 180\us-5ms                    \\ \hline
11 & \multirow{2}{*}{GPT2 Training}         & DP          & Megatron                                                            & \begin{tabular}[c]{@{}c@{}}ncclAllReduce FP16/FP32 for gradient \\ aggregation and optimizer state synchronization\end{tabular} & 300\us-20ms                   \\ \cline{1-1} \cline{3-6} 
12 &                                         & TP          & Megatron                                                            & ncclBroadcast for activation distribution                                                                                       & 300\us-6ms                    \\ \hline
\end{tabular}
\caption{The bubble summarization of the Inter-GPU Communication type.}
\end{table*}

\section{Small Bubble Detection}
\subsection{Pattern Summarization}

This section summarizes the small bubbles encountered in production frameworks and popular models across various settings. We conducted comprehensive studies on \textbf{6} models, including Qwen2.5-7B \cite{qwen2025qwen25technicalreport}, GPT2 \cite{GPT}, Llama3-8/70B \cite{Llama}, Deepseek-16B \cite{deepseek}, and GPT-oss-120B \cite{gptoss120b}, and \textbf{6} types of frameworks, including vLLM \cite{vLLM}, llama.cpp \cite{llama.cpp}, SGLang \cite{sglang}, TensorRT-LLM \cite{tensorRTLLM}, DeepSpeed \cite{deepspeed}, and Megatron \cite{megatron}. Our workloads span from inference to training. Our analysis covers diverse distributed settings, including data parallelism (DP), tensor parallelism (TP), pipeline parallelism (PP), and expert parallelism (EP). All settings use default parameters or are guided by the community to promise the best performance. Based on the root cause, we categorize these bubbles into three distinct types: 1. Memory Operation and Synchronization; 2. Inter-GPU Communication; 3. CPU-side Bound (Runtime Overhead).

\MyPara{Memory Operation and Synchronization.} This category of bubbles stems from host-device synchronizations and memory operations. This type of bubble can consume approximately 10\%-20\% of the total GPU time, depending on the model size and the frameworks used. All observed bubble patterns are summarized in Table \textcolor{DarkGreen}{\ref{tab: memory type bubble}}. First, in LLM inference, requirements for streaming responses and continuous batching enforce a strict control-flow dependency. The GPU must synchronize with the CPU at every iteration to transfer generated tokens (D2H) and receive metadata for the next batch (H2D), creating periodic small bubbles (case \textcolor{DarkGreen}{1-2}).  Second, during LLM training, the optimizer step involves a massive sequence of memory operations to combine the discontinuous gradient tensors in memory space into a contiguous buffer, reducing the number of optimize kernel calls (case \textcolor{DarkGreen}{3}). Besides, the forward phase computation relies on the input data of new batches, which need a memory copy from DRAM to HBM (case \textcolor{DarkGreen}{4}). We also observed the small bubbles in the checkpoint store phase, which involves a series of memory operations and takes about 300\us per one (case \textcolor{DarkGreen}{5}).

\MyPara{Inter-GPU Communication.} This category of bubbles stems from inter-GPU communication and synchronizations, typically accounting for more than 20\% or even 30\% of the total GPU time. All observed bubble patterns are summarized in Table \textcolor{DarkGreen}{2}.

In distributed LLM inference, we observe two major categories of bubble patterns. The first consists of CUDA APIs (\eg, \codeIn{cudaMemcpyAsync(D2D)} and \codeIn{cudaMemcpyPeerAsync}). For example, case~\textcolor{DarkGreen}{6} shows a pipeline-parallelism bubble where inference engines must copy data from communication buffers to compute buffers through a series of \codeIn{cudaMemcpyAsync(D2D)} operations. A more common bubble pattern arises from communication-library APIs (\eg, NCCL) and custom optimized communication kernels (cases~\textcolor{DarkGreen}{7-10}). Although these communication APIs ultimately execute as kernels, we still categorize them as small bubbles. This is attributed to the fact that communication kernels typically utilize a minimal fraction of SM resources (none exceeding 20\%), with only 6\% of warps in the compute state. This implies that the majority of warps remain stalled, awaiting data from remote GPUs. Such characteristics ensure that these kernels do not contend with low-priority kernels or trigger stranded-block effects. Furthermore, we can enforce stricter compute isolation via CUDA Green Context~\cite{green_ctx}. Additionally, these kernels impose negligible HBM bandwidth pressure (<0.1\% on average), relying predominantly on NVLink for data access, thereby eliminating bandwidth interference. Our experiments confirm that, during co-running, the execution times of these communication kernels exhibit almost no measurable increase, which means nearly no interference. In distributed LLM training, it is common to use NCCL APIs to perform inter-GPU communications, such as gradient aggregation (cases~\textcolor{DarkGreen}{11-12}). 


\MyPara{CPU-side Bound (Runtime Overhead).} This category of bubbles manifests when the host fails to submit kernels fast enough to keep the GPU busy, typically accounting for 5--8\% of the total GPU time. One primary cause is \textit{Module Loading}, observed in frameworks like vLLM, where lazy kernel loading or JIT compilation triggers substantial pauses during initialization. These events are characterized by \codeIn{cuModuleLoad} driver APIs or \codeIn{LazyFunctionLoading} tags. A second factor is \textit{Lock Contention} within the runtime (\eg, for the CUDA context lock or Python GIL), which can significantly delay kernel launches. This introduces launch jitter, causing the GPU command queue to drain completely before the next instruction arrives. These bubbles are identified via API calls in OS runtime libraries, such as \codeIn{pthread\_mutex\_lock}. Furthermore, CPU-side scheduling overheads, such as metadata preparation for the kernel launch, sometimes cause small bubbles.

\subsection{Automatic Pattern Discovery}
This section details our automated methodology for discovering small bubble patterns, structured as a two-phase pipeline: \textit{filtering} and \textit{verification}.

\MyPara{Phase 1: Candidate Filtering.} We first identify potential bubble patterns from the profiling logs of high-priority tasks. Our primary tool, Nsight Systems \cite{nsight-system}, provides comprehensive metrics including SM utilization, HBM bandwidth, and NVLink throughput. By exporting traces to structured formats (\eg, SQLite), we programmatically query the precise timestamps of kernel and API events. Crucially, by correlating CPU-side CUDA API traces with GPU-side hardware metrics on a unified nanosecond-precision timeline, we define 'bubbles' as intervals exhibiting zero hardware utilization despite active host-side processing (\eg, memory operations) or the execution of lightweight communication kernels (\eg, NCCL). This cross-stack visibility enables the extraction of API call patterns associated with these bubbles.

\MyPara{Phase 2: Empirical Verification.} To eliminate false positives, we verify whether these candidates can be utilized without interfering with high-priority tasks. We co-locate the high-priority tasks with compute-intensive training tasks and schedule low-priority kernels during the bubbles. If the preemption latency degradation exceeds a strict threshold (currently set to 1\%), we mark the pattern as a false positive and exclude it from deployment. The remaining patterns are validated as exploitable bubbles suitable for filling with low-priority kernels.

\section{Memory Management}
We implemented \name atop the open-source HUVM\cite{HUVM} driver stack, a derivative of the NVIDIA CUDA Unified Virtual Memory (UVM) driver. \name manages GPU physical memory at the granularity of 2MB chunks, associating each chunk with metadata that tracks its state and physical address. To facilitate efficient allocation, the system maintains a per-GPU linked list of metadata for free chunks, allowing \name to rapidly identify and harvest available spare memory of other GPUs. When a GPU borrows memory from a neighbor, the metadata is updated to reflect this cross-device usage. Furthermore, we augmented the baseline implementation with a priority isolation mechanism and replaced the default round-robin eviction logic with a contention-aware policy, driven by a runtime monitor for real-time bandwidth profiling.

\MyPara{Priority-based Memory Guidance.}
To enforce strict isolation for high-priority tasks, we leverage the CUDA Unified Memory hint API, \codeIn{cuMemAdvise}. Upon the initialization of high-priority contexts, \name automatically applies \codeIn{CU\_MEM\_ADVISE\_SET\_PREFERRED\_LOCATION} to their allocated memory regions, pinning them to the local GPU. This hint instructs the driver to skip the pages marked as "preferred", preventing the eviction of high-priority tasks' data under memory pressure. When memory contention occurs, the driver's page fault handler checks these advice flags. The pages belonging to high-priority tasks are skipped during victim selection.

\MyPara{Interference-Aware Eviction Policy.} We modified the page eviction policy in the HUVM driver to replace the original round-robin eviction policy. We redesigned this mechanism to select the optimal eviction destination based on real-time link congestion. Specifically, we maintain a dynamic \textit{congestion score} table for all NVLink connections. During the victim selection phase, instead of selecting the next peer in a circular order, the eviction thread queries this table to identify the peer GPU with the lowest congestion score. If the target peer GPU has free memory and its link congestion is below a predefined threshold (indicating interference-free bandwidth), the page is migrated there via NVLink. If all peer GPUs are either saturated in memory or experiencing high bandwidth contention (high latency), the policy falls back to the host path, evicting the page to system DRAM. This \textit{contention-first} strategy effectively prevents low-priority memory swap from interfering with local high-priority kernel execution.

\MyPara{Real-time Bandwidth Monitoring.}
To maintain the congestion score table, \name deploys a lightweight global monitor running as a background daemon. This monitor employs a periodic \textbf{ping-like} probing mechanism to measure NVLink latency without disrupting active workloads, which includes the following three key components:
\begin{itemize}
\item \textit{Baseline Calibration:} At system startup, the monitor performs a calibration phase by transferring a fixed-size probe packet (e.g., 4MB) between all GPU pairs to establish a baseline latency ($T_{base}$) under idle conditions.
\item \textit{Periodic Probing:} During runtime, the monitor periodically sends the same probe packet and measures the current latency ($T_{curr}$). The congestion score is calculated as the ratio $T_{curr} / T_{base}$. A significant deviation (e.g., $>1.5$) signals heavy bandwidth contention, likely caused by concurrent kernel execution or collective communications.
\item \textit{Low-Overhead Design:} The probing frequency is adaptively adjusted (typically 10ms--100ms) to ensure the monitoring overhead remains negligible ($<0.1\%$ of total bandwidth consumption). The collected scores are written to a shared memory region mapped into the driver's address space, allowing the eviction policy to make decisions with sub-microsecond latency.
\end{itemize}

\end{document}